\documentclass{iopart} 
\usepackage{enumerate}
\usepackage[T1]{fontenc}
\expandafter\let\csname equation*\endcsname\relax
\expandafter\let\csname endequation*\endcsname\relax
\usepackage{amsmath, amsthm, amssymb, amsfonts}
\usepackage{graphicx}
\usepackage[colorlinks=true,linkcolor=blue,citecolor=blue,filecolor=cyan,urlcolor=blue,breaklinks=false]{hyperref}
\usepackage{cite}
\usepackage[ugly]{units}
\usepackage{booktabs}

\newcommand{\Ftot}{F}

\newcommand{\TOexp}[1]{\overrightarrow{\exp}\left({#1}\right)}
\newcommand{\T}{\mathcal{T}}
\newcommand{\expval}[1]{\langle #1 \rangle}

\newcommand{\pspeed}{{v}}

\newcommand{\DDelta}{\Delta}
\newcommand{\dd}[1]{\mathrm{d}#1\,}
\newcommand{\order}[1]{\mathcal{O}\left(#1\right)}

\newcommand{\Aif}{a_\T}
\newcommand{\Aim}{a(\T,\pspeed)}
\newcommand{\Aimorder}[1]{a^{(#1)}(\T,\pspeed)}
\newcommand{\Acf}{A_\T}
\newcommand{\Acm}{A(\T,\pspeed)}
\newcommand{\Acmorder}[1]{A^{(#1)}(\T,\pspeed)}
\newcommand{\Jf}{J_\T}
\newcommand{\Jm}{J(\T,\pspeed)}
\newcommand{\Jtf}[1]{J^{#1}_\T}
\newcommand{\Jtm}[1]{J_{#1}(\T,\pspeed)}
\newcommand{\Jtmorder}[2]{J^{(#1)}_{#2}(\T,\pspeed)}

\newcommand{\statevar}{\mathcal{A}}
\newcommand{\statevarf}{\statevar_\T}
\newcommand{\statevarm}{\statevar(\T,\pspeed)}

\newcommand{\tsys}{t_\mathrm{sys}}

\newcommand{\tf}{\tau_f}

\newcommand{\mobility}{\mu}
\newcommand{\D}{D}
\newcommand{\scaledsymbol}[1]{\tilde{#1}}
\newcommand{\smobility}{\scaledsymbol{\mu}}
\newcommand{\sD}{\scaledsymbol{D}}
\newcommand{\sdensity}{\scaledsymbol{p}}
\newcommand{\sj}{\scaledsymbol{j}}

\newcommand{\FPO}[1][x,\lambda_{\tau}]{\mathcal{L}_\mathrm{FP}\left(#1\right)}
\newcommand{\FPOscaled}[1][x,\lambda_{\tau}]{\hat{\mathcal{L}}_\mathrm{FP}\left(#1\right)}
\newcommand{\FPOeff}[1][x]{\hat{\mathcal{L}}_\mathrm{eff}\left(#1\right)}
\newcommand{\TEO}[3]{\hat{U}\left({#1},{#2},{#3}\right)}
\newcommand{\TEOunscaled}[3]{U\left({#1},{#2},{#3}\right)}
\newcommand{\TEOorder}[4]{\hat{U}^{(#1)}\left({#2},{#3},{#4}\right)}

\newcommand{\epsf}{\epsilon_f}
\newcommand{\epss}{\epsilon_s}
\newcommand{\porder}[1]{p^{(#1)}}
\newcommand{\jorder}[1]{j^{(#1)}}

\newcommand{\response}[1]{\mathcal{R}_{#1}(\T,\pspeed)}

\begin{document}
\title[]{Quality of the Thermodynamic Uncertainty Relation for Fast and Slow Driving}
\author{Timur Koyuk$^1$ and Udo Seifert$^1$}

\address{$^1$ II. Institut f\"ur Theoretische Physik, Universit\"at Stuttgart, 70550 Stuttgart, Germany}

\begin{abstract}
  The thermodynamic uncertainty relation originally proven for systems driven
  into a non-equilibrium steady state (NESS) allows one to infer the total entropy production rate by observing
  \textit{any} current in the system. This kind of inference scheme is especially useful when the
  system contains hidden degrees of freedom or hidden discrete states, which
  are not accessible to the experimentalist. A recent generalization of the
  thermodynamic uncertainty relation to arbitrary time-dependent driving
  allows one to infer entropy production not only by measuring current-observables
  but also by observing \textit{state variables}. A crucial question then is
  to understand which observable
  yields the best estimate for the total entropy production. In this paper we
  address this question by analyzing the quality of the thermodynamic
  uncertainty relation for various types of observables for the generic
  limiting cases of fast driving and slow driving. We show that in both
  cases observables can be found that yield an estimate of order one for
  the total entropy production. We further show that the uncertainty
  relation can even be saturated in the limit of fast driving.
  \\[1\baselineskip]
  \noindent{\it Dated\/}: \today
\end{abstract}

\section{Introduction}
Recent progresses in the field of non-equilibrium statistical physics have reshaped
our perspective on conventional thermodynamic notions such as work, heat or
entropy production. Defining these thermodynamic observables along single
fluctuating trajectories is the key step to build up a
theoretical formalism nowadays called \textit{stochastic
  thermodynamics}~\cite{seki10,jarz11,seif12,vdb15}. As a key property of these small mesoscopic systems
fluctuations and their relation to universal non-equilibrium properties are of
special interest from a theoretical as well as from an operational or
experimental point of view.
A well-established paradigm for such a connection is
the \textit{fluctuation-dissipation theorem} (FDT) relating
equilibrium fluctuations to the dissipation rate in driven systems near
equilibrium~\cite{kubo66}. Milestones in the field of stochastic
thermodynamics \textit{inter alia} deal with similar connections for systems \textit{far
away} from equilibrium: from fluctuation
theorems~\cite{evan93,gall95,kurc98,lebo99,evan94,jarz97,jarz97a,croo99,croo00,seif05}
and generalizations of the FDT~\cite{marc08,baie09,pros09,seif09,baie12} to the Harada-Sasa relation connecting the
violation of the FDT to energy dissipation~\cite{hara05,hara06}.

A more recent development in this lineup is the so-called
\textit{thermodynamic uncertainty relation} (TUR), which connects the
fluctuations or precision of \textit{any} current in the system to the total
entropy production rate~\cite{bara15,ging16}. For an overdamped Langevin
system or a Markovian system on a discrete set of states driven into a NESS
the thermodynamic uncertainty relation for finite observation times $\T$ reads~\cite{piet17,horo17}
\begin{equation}
  \label{eq:intro_TUR_NESS}
  D_J(\T)\sigma(\T)/J(\T)^2 \ge 1
\end{equation}
with current $J(\T)$, its diffusion coefficient $D_J(\T)$ quantifying fluctuations
and the total entropy production rate $\sigma(\T)$. Beyond considering the TUR as
a trade-off relation between precision and dissipation leading to bounds on
the efficiency of biological processes or molecular
machines~\cite{piet16b,piet17a,hwan18} it has been established as a useful tool for
inferring entropy production~\cite{ging16a,li19,dech20,mani20}. Hence, numerous attempts have been made to extend the
range of applicability of the TUR including underdamped
dynamics~\cite{fisc18,dech18,chun19,lee19,fisc19}, ballistic transport between
different terminals~\cite{bran18}, heat
engines~\cite{shir16,piet17a,holu18,ekeh20}, periodic driving~\cite{proe17,bara18b,koyu19,proe19,koyu19a}, stochastic field
theories~\cite{nigg20,nigg21}, generalizations to observables that are even under
time-reversal~\cite{maes17,nard17a,terl18}, first-passage time problems~\cite{ging17,garr17,hiu21} and
quantum
systems~\cite{maci18,agar18,ptas18,bran18,carr19,guar19,caro19,pal20,frie20}.

In this vast lineup of generalizations and ramifications of the TUR each
relation has its own region of validity. A unifying uncertainty relation for
arbitrary time-dependent driving including the TUR for finite observation times~\cite{piet17,horo17}, for relaxation
processes~\cite{dech17,liu19} and for periodically driven
systems~\cite{koyu19a} has been found recently~\cite{koyu20}. This relation reads
\begin{equation}
  \label{eq:intro_TUR_TDD}
  D_J(\T,\pspeed)\sigma(\T,\pspeed)/J(\T,\pspeed)^2 \ge \left[1 + \DDelta J(\T,\pspeed)/J(\T,\pspeed)\right]^2,
\end{equation}
where the speed of driving
$\pspeed$ enters as the second argument and $\DDelta J(\T,\pspeed)$ describes the
change of the current with respect to the observation time $\T$ and the speed
of driving $\pspeed$. A similar inequality involving the total entropy
production rate can also be derived for state
variables~\cite{koyu20}. Since their origin lies in the response of the system with respect
to a time re-scaling by using a virtual perturbing force~\cite{dech18a}, these
relations should be clearly distinguished from so-called generalized
thermodynamic uncertainty relations that are solely a consequence of the
fluctuation theorem~\cite{hase19,timp19}. Furthermore, the TUR for
time-dependent driving~\eqref{eq:intro_TUR_TDD} involves operationally accessible
observables and hence, preserves the desired property of
being a trade-off relation between those. It thus remains a useful tool for inferring entropy
production, in principle. However, the question remains, which observables yield the best estimate for entropy production. 

In this paper, we analyze the quality of the thermodynamic uncertainty
relation (TUR) for time-dependent driving~\eqref{eq:intro_TUR_TDD} in the limiting cases of
fast driving and slow driving for overdamped Langevin systems. We show that in
each limiting case at least one optimal class of observable
exists that generically yields an estimate of order one for the total entropy
production rate. We further show that the time-dependent uncertainty relation in
ref.~\cite{koyu20} simplifies to the conventional form of the steady-state TUR
in refs.~\cite{bara15,ging16} in the fast-driving limit. We demonstrate that
in this limiting case a current proportional to the total entropy production
rate can saturate the TUR. For the slow-driving limit we show that the choice of the optimal
observable depends on whether or not a non-conservative force is applied.
Moreover, we show that these results hold not only for systems
with continuous degrees of freedom, but also for systems with a discrete set
of states as we illustrate for a driven three-state model.

\section{Setup}
\subsection{Dynamics}
We consider a system with one continuous degree of freedom $x(t)$. The dynamics
is given by an overdamped Langevin equation
\begin{equation}
  \label{eq:setup_langevin_eq}
  \dot{x}(t) = \mu  \Ftot(x(t),\lambda_t) + \zeta(t),
\end{equation}
where $\zeta(t)$ is a zero-mean Gaussian white noise satisfying
\begin{align}
  \label{eq:setup_noise_relation_1}
  \expval{\zeta(t)} &= 0, \\
  \label{eq:setup_noise_relation_1}
  \expval{\zeta(t)\zeta(t')} &= 2D\delta(t-t').
\end{align}
The system is driven by a time-dependent force
\begin{equation}
  \label{eq:setup_total_force}
  \Ftot(x,\lambda_t) \equiv f(\lambda_t) - \partial_x V(x,\lambda_t),
\end{equation}
which consists of a non-conservative force $f(\lambda_t)$ and a conservative
part $-\partial_x V(x,\lambda_t)$. Both contributions depend on a
time-dependent protocol $\lambda_t \equiv \lambda(\pspeed t)$.
Here, $\pspeed$ denotes the speed of driving and $D\equiv \mobility/\beta$ is the diffusion constant,
where $\mobility$ is the mobility and $\beta$ is the inverse temperature.
Equivalently, we can use a Fokker-Planck equation
\begin{equation}
  \label{eq:derivation:Fokker-Planck-Equation}
  \partial_t p(x,t) = - \partial_x (\mobility \Ftot(x,\lambda_t) - D \partial_x) p(x,t)
\end{equation}
describing the dynamics for the probability $p(x,t)$
to find the system in state $x$ at time $t$. The system is
observed up to time $\T$, where the protocol $\lambda_t$ evolves from value
$\lambda(0)$ to $\lambda(\tf\equiv\pspeed\T)$. In the following, we keep the
final value $\tf$ of the protocol fixed, i.e., the observation time
$\T=\tf/\pspeed$ is coupled to the speed of
driving. Equation~\eqref{eq:derivation:Fokker-Planck-Equation} describes
probability conservation and hence, is a continuity equation for the probability
current
\begin{equation}
  \label{eq:derivation_probabilty_current}
  j(x,t) \equiv (\mobility \Ftot(x,\lambda_t) - D\partial_x) p(x,t).
\end{equation}

\subsection{Observables}
The framework of stochastic thermodynamics allows us to define several types
of observables for arbitrary time-dependent driven
systems~\cite{seif12,koyu20}.
These observables depend on the state $x(t)$ of the system or the velocity
$\dot{x}(t)\equiv\partial_t x(t)$. The first type of
observable we are focusing on is called a state variable $a(x,\lambda_t)$. This variable can be
either observed at a fixed observation time
\begin{equation}
  \label{eq:derivation_instant_state_var}
  \Aif \equiv a(x_\T,\lambda_\T)
\end{equation}
or it can be time-averaged over a finite-time $\T$
\begin{equation}
  \label{eq:derivation_time_averaged_state_var}
  \Acf \equiv \frac{1}{\T}\int_0^\T\dd{t} a(x_t,\lambda_t),
\end{equation}
where $x_t \equiv x(t)$.
The second kind of observable is a current, which is odd under time
reversal. Here, we distinguish between a current depending on the time spent
in a certain state
\begin{equation}
  \label{eq:derivation_current_I}
  \Jtf{b} \equiv \frac{1}{\T}\int_0^\T\dd{t} \dot{b}(x_t,\lambda_t),
\end{equation}
which depends on the time-derivative of a state variable
$\dot{b}(x_t,\lambda_t)\equiv (\partial_t\lambda_t)\partial_{\lambda}b(x_t,\lambda)\vert_{\lambda=\lambda_t}$ and a current depending on the velocity, i.e.,
\begin{equation}
  \label{eq:derivation_current_II}
  \Jtf{d} \equiv \frac{1}{\T}\int_0^\T\dd{t} d(x_t,\lambda_t)\circ\dot{x}_t,
\end{equation}
where $d(x_t,\lambda_t)$ is a function of the state and $\circ$ denotes the
Stratonovich product. A further important observable of interest is the mean total
entropy production rate 
\begin{equation}
  \label{eq:derivation_total_entropy_production_rate}
  \sigma(\T,\pspeed) \equiv
  \frac{1}{\T}\int_0^{\T}\dd{t}\int\dd{x}\frac{j^2(x,t)}{D p(x,t)},
\end{equation}
The fluctuations around the mean
value $\expval{X_\T}$ of any of the above introduced observables $X_\T\in\{\Aif,\Acf, \Jtf{b,d}\}$ are quantified by the diffusion coefficient
\begin{equation}
  \label{eq:derivation_diffusion_coefficient}
D_X(\T,\pspeed) \equiv \T\left(\expval{X_\T^2} - \expval{X_\T}^2\right)/2,
\end{equation}
where $\expval{\cdot}$ denotes the mean value.

\subsection{Quality Factors and the Thermodynamic Uncertainty Relation}
The recent generalization of the thermodynamic uncertainty relation to
arbitrary time-dependent driving~\cite{koyu20} can be applied to all types of
observables defined in
eqs.~\eqref{eq:derivation_instant_state_var}--\eqref{eq:derivation_current_II}.
For current-type observables
$\Jf\in\{\Jtf{d},\Jtf{b}\}$ the uncertainty relation
\begin{equation}
  \label{eq:setup_quality_factors_UCR_current}
  1 \ge \frac{\response{J}}{D_{J}(\T,\pspeed)\sigma(\T,\pspeed)}
\end{equation}
imposes a bound in terms of the response term
\begin{equation}
  \label{eq:setup_quality_factors_response_term_current}
  \response{J} \equiv [\Jm + \DDelta \Jm]^2
\end{equation}
with mean value $\Jm\equiv\expval{\Jf}$ and operator $\DDelta \equiv \T \partial_\T - \pspeed
\partial_\pspeed$. The term $\DDelta \Jm$ describes the change of the
current with respect to a slight change of the observation time $\T$ and the
speed of driving $\pspeed$. For state variables $\statevarf\in\{\Aif,\Acf\}$
the uncertainty relation
\begin{equation}
  \label{eq:setup_quality_factors_UCR_state_variable}
  1 \ge \frac{\response{\statevar}}{D_{\statevar}(\T,\pspeed)\sigma(\T,\pspeed)}
\end{equation}
involves a modified response term
\begin{equation}
  \label{eq:setup_quality_factors_response_term_state_variable}
  \response{\statevar} \equiv [\DDelta \statevarm]^2,
\end{equation}
where $\statevar(\T,\pspeed)\equiv\expval{\statevarf}$ denotes the mean value of a state
variable.

For both relations eqs.~\eqref{eq:setup_quality_factors_UCR_current}
and~\eqref{eq:setup_quality_factors_UCR_state_variable} we define the quality
factors as
\begin{equation}
  \label{eq:setup_quality_factors_definition_quality_factors_current}
  \mathcal{Q}_J \equiv \frac{\response{J}}{D_{J}(\T,\pspeed)\sigma(\T,\pspeed)}
\end{equation}
and
\begin{equation}
  \label{eq:setup_quality_factors_definition_quality_factors_state_variable}
  \mathcal{Q}_\statevar \equiv \frac{\response{\statevar}}{D_{\statevar}(\T,\pspeed)\sigma(\T,\pspeed)},
\end{equation}
respectively. Both quality factors are always larger than zero and smaller
than one. If a quality factor is zero, no information can be inferred about the
entropy production by observing the response and fluctuations of an
observable. However, if a quality factor is one, the uncertainty relation is
saturated and we can determine the total entropy production exactly.

\subsection{Time Scale Separation}
\label{sec:time_scale_separation}
The aim of this paper is to analyze the limits of fast driving and slow driving.
Hence, it is useful to introduce a time-scale separation between the time scale of the
system $\tsys$ and the time scale of the driving $\pspeed^{-1}$. If the relaxation time scales in the system are approximately
of the same order of magnitude, we can choose the basic time scale of the system
$\tsys$ as this order of magnitude. However, if the
relaxation time
scales have different orders of magnitudes, e.g., due to a complex
topology like energy barriers
in the system, we have to distinguish between the
limiting cases of fast driving and slow driving: for the fast-driving limit the basic time scale of the system $\tsys$ has to be
chosen as the relaxation time scale describing the fastest relaxation. In contrast, for the
limit of slow driving we have to chose $\tsys$ as the time scale that
describes the slowest relaxation.
Depending on the above discussed cases, the fastest or slowest relaxation time
scale in the system is proportional to the inverse of the mobility $\mobility$. Hence, the mobility is proportional to the inverse of
the time scale $\tsys$ of the
system. This circumstance allows us to define a scaled mobility
\begin{equation}
  \label{eq:derivation_scaled_mobility}
  \smobility \equiv \mobility \tsys.
\end{equation}
Plugging eq.~\eqref{eq:derivation_scaled_mobility}
into eq.~\eqref{eq:derivation:Fokker-Planck-Equation} and using the
substitution $\tau\equiv \pspeed t = \tf t/\T$ leads to the scaled Fokker-Planck equation
\begin{equation}
  \label{eq:derivation_scaled_FPE}
  \partial_\tau \sdensity(x, \tau) = -\left(\frac{\T}{\tf\tsys}\right)\partial_x (\smobility \Ftot(x,\lambda_\tau) - \sD \partial_x) \sdensity(x,\tau)
\end{equation}
with a scaled diffusion constant $\sD \equiv \smobility/\beta$. The density $\sdensity(x,\tau)
\equiv p(x,\tau/\pspeed)$ depends on the speed of driving $\pspeed$, i.e.,
\begin{equation}
  \label{eq:derivation:p_as_function_of_mu_and_v}
  \sdensity(x,\tau) = \sdensity(x,\tau;\pspeed).
\end{equation}
We further define the scaled probability current as
\begin{equation}
  \label{eq:derivation_scaled_probabilty_current}
  \sj(x,\tau;\pspeed) \equiv (\smobility \Ftot(x,\lambda_\tau) - \sD
  \partial_x) \sdensity(x,\tau;\pspeed).
\end{equation}

For the sake of simplicity, we change the notation
$\sdensity(x,\tau;\pspeed) \to p(x,\tau;\pspeed)$ and $\sj(x,\tau;\pspeed)\to
j(x,\tau;\pspeed)$ in the following. The scaled Fokker-Planck
equation~\eqref{eq:derivation_scaled_FPE} then reads 
\begin{equation}
  \label{eq:derivation_scaled_FPE_new_notation}
  \partial_\tau p(x, \tau;\pspeed) = (\pspeed\tsys)^{-1} \FPOscaled p(x,\tau;\pspeed),
\end{equation}
with the scaled Fokker-Planck operator
\begin{equation}
  \label{eq:derivation_scaled_FPO}
  \FPOscaled \equiv -\partial_x(\smobility \Ftot(x,\lambda_\tau) - \sD \partial_x).
\end{equation}
The general solution of eq.~\eqref{eq:derivation_scaled_FPE_new_notation}
for a given initial distribution $p(x,0)$ reads
\begin{equation}
  \label{eq:derivation_general_solution}
  p(x,\tau;\pspeed) = \TEO{x}{\tau}{0} p(x,0),
\end{equation}
where 
\begin{equation}
  \label{eq:derivation_time_evolultion_operator}
  \TEO{x}{\tau_2}{\tau_1} \equiv \TOexp{\int_{\tau_1}^{\tau_2}\dd{\tau}(\pspeed\tsys)^{-1}\FPOscaled}
\end{equation}
is the time evolution operator and $\TOexp{\cdot}$ denotes a time-ordered
exponential. Via eq.~\eqref{eq:derivation_time_evolultion_operator} we can define the
propagator as
\begin{equation}
  \label{eq:derivation_propagator}
  p(x_2,\tau_2\vert x_1,\tau_1) \equiv \TEO{x_2}{\tau_2}{\tau_1}\delta(x_2 - x_1),
\end{equation}
where $\tau_2 \ge \tau_1$ and $\delta(\cdot)$ denotes a Dirac delta.

The mean values of the state variables~\eqref{eq:derivation_instant_state_var}
and~\eqref{eq:derivation_time_averaged_state_var} in terms of the scaled time
$\tau=\pspeed t = \tf t/\T$ are given by
\begin{equation}
  \label{eq:derivation_mean_value_instant_state_var}
  \Aim \equiv \int\dd{x} a(x,\lambda_{\tf})p(x,\tf;\pspeed)
\end{equation}
and
\begin{equation}
  \label{eq:derivation_mean_value_time_averaged_state_var}
  \Acm \equiv \frac{1}{\tf}\int_0^{\tf}\dd{\tau}\int\dd{x}a(x,\lambda_{\tau})p(x,\tau;\pspeed),
\end{equation}
respectively, whereas the mean values of the currents~\eqref{eq:derivation_current_I}
and~\eqref{eq:derivation_current_II} are given by
\begin{equation}
  \label{eq:derivation_mean_value_current_I}
  \Jtm{b} \equiv \frac{v}{\tf}\int_0^{\tf}\dd{\tau}\int\dd{x}\dot{b}(x,\lambda_{\tau})p(x,\tau;\pspeed)
\end{equation}
and
\begin{equation}
  \label{eq:derivation_mean_value_current_II}
  \Jtm{d} \equiv \frac{1}{\tsys\tf}\int_0^{\tf}\dd{\tau}\int\dd{x}d(x,\lambda_{\tau})j(x,\tau;\pspeed),
\end{equation}
respectively.
Here, $\dot{b}(x,\lambda_{\tau}) \equiv \partial_\tau b(x,\lambda_{\tau})$ is
the time-derivative in terms of time scale $\tau$ and
\begin{equation}
  \label{eq:derivation_probability_current}
  j(x,\tau;\pspeed) \equiv (\smobility \Ftot(x,\lambda_\tau) - \sD\partial_x)p(x,\tau;\pspeed)
\end{equation}
is the scaled probability current.
Moreover, the mean total entropy production rate~\eqref{eq:derivation_total_entropy_production_rate} in terms of the scaled
quantities is given by
\begin{equation}
  \label{eq:derivation_scaled_total_entropy_production_rate}
  \sigma(\T,\pspeed) \equiv
  \frac{1}{\tsys\tf}\int_0^{\tf}\dd{\tau}\int\dd{x}\frac{j^2(x,\tau;\pspeed)}{\sD p(x,\tau;\pspeed)}.
\end{equation}
The diffusion coefficients of the quantities defined in
eqs.~\eqref{eq:derivation_instant_state_var}--\eqref{eq:derivation_current_II}
can be written in terms of correlation functions between state variables and
hence, depend on the propagator~\eqref{eq:derivation_propagator}. Their explicit
expressions in terms of the scaled time $\tau$
can be found in~\ref{sec:appendix_diffusion_coefficients}.

\section{Fast Driving}
\label{sec:fast_driving}
We first consider the limit of fast driving, where the driving is much faster
than the fastest relaxation time scale of the system.
The limit of fast driving requires the parameter
\begin{equation}
  \label{eq:derivation_fast_driving_eps1}
  \epsf \equiv \frac{1}{\pspeed\tsys} \ll 1
\end{equation}
to be small, i.e., $\pspeed^{-1} \ll
\tsys$ or equivalently, $\T\ll \tf\tsys$. This means that the time scale of the driving $\pspeed^{-1}=\T/\tf$ is much
shorter than the time scale $\tsys$ on which the fastest relaxation of the system takes
place. The time evolution operator in
eq.~\eqref{eq:derivation_time_evolultion_operator} can be expanded in terms of
$\epsf$, i.e.,
\begin{equation}
  \label{eq:derivation_fast_driving_time_evolution_operator_expansion}
  \TEO{x}{\tau_2}{\tau_1} = 1 + \epsf \int_{\tau_1}^{\tau_2}\dd{\tau}
  \FPOscaled + \order{\epsf^2}.
\end{equation}
Via eq.~\eqref{eq:derivation_general_solution} the density is given by
\begin{equation}
  \label{eq:derivation_fast_driving_density}
  p(x,\tau;\pspeed) = \porder{0}(x,\tau) + \epsf
  \porder{1}(x,\tau) + \order{\epsf^2}
\end{equation}
with zeroth and first order
\begin{align}
  \label{eq:derivation_fast_driving_density_zero_order}
  \porder{0}(x,\tau) &= p(x,0), \\
  \label{eq:derivation_fast_driving_density_first_order}
  \porder{1}(x,\tau) &= \FPOeff[x,\tau,0] p(x,0),
\end{align}
respectively, and
\begin{equation}
  \label{eq:derivation_fast_driving_time_averaged_FPO}
  \FPOeff[x,\tau,0] \equiv \int_{0}^{\tau}\dd{\tau'} \FPOscaled[x,\lambda_{\tau'}]
\end{equation}
being the time-averaged Fokker-Planck operator. The probability current is
analogously given by
\begin{equation}
  \label{eq:derivation_fast_driving_probabilty_current}
  j(x,\tau;\pspeed) = \jorder{0}(x,\tau) + \epsf
  \jorder{1}(x,\tau) + \order{\epsf^2}
\end{equation}
with zeroth and first order
\begin{align}
  \label{eq:derivation_fast_driving_current_zero_order}
  \jorder{0}(x,\tau) &= (\smobility \Ftot(x,\lambda_\tau) - \sD \partial_x)p(x,0), \\
  \label{eq:derivation_fast_driving_current_first_order}
  \jorder{1}(x,\tau) &= (\smobility \Ftot(x,\lambda_\tau) - \sD \partial_x)\FPOeff[x,\tau,0] p(x,0),
\end{align}
respectively. The leading order of the
density~\eqref{eq:derivation_fast_driving_density_zero_order} shows that the
fast driving leaves the initial distribution over the observation time
unchanged. The density can then approximately be described by the
time-independent initial condition. As a consequence, the leading order of the
probability current~\eqref{eq:derivation_fast_driving_current_zero_order}
depends only on the protocol.
Furthermore, we can
use~\eqref{eq:derivation_fast_driving_time_evolution_operator_expansion} to
get the leading orders of the propagator~\eqref{eq:derivation_propagator},
i.e.,
\begin{equation}
  \label{eq:derivation_fast_driving_propagator}
  p(x_2,\tau_2\vert x_1,\tau_1) = \porder{0}(x_2,\tau_2\vert x_1,\tau_1) +
  \epsf \porder{1}(x_2,\tau_2\vert x_1,\tau_1) + \order{\epsf^2}.
\end{equation}
with zeroth and first order
\begin{align}
  \label{eq:derivation_fast_driving_propagator_zero_order}
  \porder{0}(x_2,\tau_2\vert x_1,\tau_1) &= \delta(x_2 - x_1),\\
  \label{eq:derivation_fast_driving_propagator_first_order}
  \porder{1}(x_2,\tau_2\vert x_1,\tau_1) &= \FPOeff[x_2,\tau_2,\tau_1] \delta(x_2 - x_1),
\end{align}
respectively.

To determine the leading orders of the quality factors for the different types
of observables, we use
eqs.~\eqref{eq:derivation_fast_driving_density},~\eqref{eq:derivation_fast_driving_probabilty_current}
as well as~\eqref{eq:derivation_fast_driving_propagator} to determine the
leading orders of the scaled mean values,
eqs.~\eqref{eq:derivation_mean_value_instant_state_var}--\eqref{eq:derivation_mean_value_current_II},
their response
terms, their corresponding diffusion coefficients,
eqs.~\eqref{eq:derivation_diff_coeff_value_instant_state_var}--\eqref{eq:derivation_diffusion_coefficient_current_II}, as well as the
scaled total entropy production
rate~\eqref{eq:derivation_scaled_total_entropy_production_rate}.
Here, all mean values and diffusion coefficients can be written as
\begin{equation}
  \label{eq:derivation_fast_driving_general_scaling_mean}
  X(\T,\pspeed) \equiv \sum_{n=0}\left(\epsf\right)^n X^{(n)}(\T,\pspeed)
\end{equation}
and
\begin{equation}
  \label{eq:derivation_fast_driving_general_scaling_diff_coeff}
  D_X(\T,\pspeed) \equiv \sum_{n=0} \left(\epsf\right)^n D_X^{(n)}(\T,\pspeed),
\end{equation}
respectively with mean values $X(\T,\pspeed)\in\{\Aim,\Acm,\epsf\Jtm{b},\Jtm{d}\}$ and
diffusion coefficients
$D_X(\T,\pspeed)\in\{D_a(\T,\pspeed), D_A(\T,\pspeed),
\epsf D_{J_{b}}(\T,\pspeed), D_{J_{d}}(\T,\pspeed) \}$.
Their leading orders and the resulting quality factors are shown in
table~\ref{tab:ATDD_Fast_Driving}
(see~\ref{sec:appendix_limit_of_fast_driving} for details of the derivation).
\begin{table}[tbp]
  \centering
  \caption{Leading orders of the total entropy production rate, the response
    terms, the diffusion coefficents and the quality factors of the observables in the limit of fast
    driving.}
  \label{tab:ATDD_Fast_Driving}
  \begin{tabular}{lllll}
    \toprule
    Observable $X$ & Response Term $\mathcal{R}_X$ & $D_X$ &
                                                           $\sigma(\T,\pspeed)$
    & $\mathcal{Q}_X$\\ 
    \midrule
    \midrule
    $X=\Aim$ & $[\DDelta \Aim]^2 = \order{\epsf^2}$ & $\order{\epsf}$
                                       & $\order{1}$ & $\order{\epsf}$\\
    \midrule
    $X=\Acm$ & $[\DDelta \Acm]^2=\order{\epsf^2}$ & $\order{\epsf}$ & $\order{1}$
                                                              &
                                                                $\order{\epsf}$\\
    \midrule
    $X=\Jtm{b}$ & $[\Jtm{b} + \DDelta \Jtm{b}]^2=\order{1}$ &
                                                        $\order{\epsf^{-1}}$
                                       & $\order{1}$ & $\order{\epsf}$\\
    \midrule
    $X=\Jtm{d}$ & $[\Jtm{d} + \DDelta \Jtm{d}]^2=\order{1}$ &
                                                        $\order{1}$
                                       & $\order{1}$ & $\order{1}$\\
    \bottomrule
  \end{tabular}
\end{table}
The response terms of the state variables vanish like $\epsf^2$, whereas
the response terms of the current observables are of $\order{1}$. The
diffusion coefficients of the state variables vanish like $\epsf$
because their variances are of $\order{1}$. This circumstance is a consequence
of the fact that the fast driving conserves the initial distribution. In
contrast to the state variables $\Aim$ and $\Acm$ the diffusion coefficient for currents depending
on the residence time $D_{J_b}$ diverges proportional to
$\epsf^{-1}$ due to the additional time-derivative of the
increment. Together with the fact that the mean total entropy
production rate is of $\order{1}$ these results imply 
quality factors that vanish linearly with $\epsf$ for all observables except for the quality factor
$\mathcal{Q}_{J_d}$, which is of $\order{1}$. 

To summarize, in the limit of fast driving generically only the current observable
$\Jtm{b}$ yields an useful estimate for entropy production.
Moreover, the explicit expression for quality factor reads (see~\ref{sec:appendix_limit_of_fast_driving})
\begin{equation}
  \label{eq:derivation_fast_driving_quality_factor_II_explicit}
  Q_{J_d} \approx
  \frac{\left[\int_0^{\tf}\dd{\tau}\int\dd{x}d(x,\lambda_\tau)\jorder{0}(x,\tau)\right]^2}{\left[\int_0^{\tf}\dd{\tau}\int\dd{x}d^2(x,\lambda_\tau)p(x,0)\right]
  \left[\int_0^{\tf}\dd{\tau}\int\dd{x} \left(\jorder{0}(x,\tau)\right)^2/(\sD
    p(x,0))\right]}.
\end{equation}
Here, the response of the current $\DDelta \Jtm{d}$ vanishes, which implies that eq.~\eqref{eq:setup_quality_factors_UCR_current} simplifies to the
conventional form of the steady-state uncertainty relation in refs.~\cite{bara15,ging16}.
Furthermore eq.~\eqref{eq:derivation_fast_driving_quality_factor_II_explicit} shows that the TUR for time-dependent driving can be saturated for the
choice
\begin{equation}
  \label{eq:derivation_fast_driving_choice_saturation}
  d(x,\lambda_\tau) = (\smobility \Ftot(x,\lambda_\tau) - D [\partial_x
  p(x,0)]/p(x,0))/\sD = \jorder{0}(x,\tau)/(\sD p(x,0)),
\end{equation}
i.e., when the current is chosen to be the total entropy production rate. We
remark that choosing the total entropy production as a current in
eq.~\eqref{eq:setup_quality_factors_UCR_current} is in general not allowed
due to the fact that the increment for the entropy production $d_\sigma \equiv j(x,\tau;\pspeed)/p(x,\tau;\pspeed)$ is
not a function of the protocol, i.e., $d_\sigma \neq d_\sigma(x,\lambda_\tau)$
(see derivation in ref.~\cite{koyu20}).
However, in the fast-driving limit the probability current becomes a function of the
protocol, i.e., $j(x,\tau;\pspeed) = \jorder{0}(x,\lambda_\tau) + \order{\epsf}$ and thus the
total entropy production rate fulfills the uncertainty
relation~\eqref{eq:setup_quality_factors_UCR_current}.
The fact that the total entropy production can always saturate the TUR in the
fast-driving limit is unique for systems with continuous degrees of
freedom. For systems with discrete degrees of freedom the definition of the
total entropy production rate prevents the saturation of the TUR arbitrary far away
from equilibrium. Only for discrete systems close to equilibrium the TUR can be
saturated~\cite{ging16,ging16a}.
Moreover, for a constant protocol the result in
eq.~\eqref{eq:derivation_fast_driving_quality_factor_II_explicit} for fast
driving reduces to the result for steady-states in refs.~\cite{mani20,dech20}
in the limit of short observation times.
As a consequence, we have generalized
this result to arbitrary time-dependent driving and shown that the total
entropy production rate can always saturate the TUR in the short-time limit
beyond steady-states for arbitrary driving.

\section{Slow Driving}
\label{sec:slow_driving}
As the second limiting case we consider the limit of slow driving, where the
time-dependent driving is much slower than the slowest relaxation time of the system.
In this limit the parameter
\begin{equation}
  \label{eq:derivation_slow_driving_eps_s}
  \epss \equiv \pspeed\tsys \ll 1
\end{equation}
is assumed to be small. Here, the time scale of the driving $\pspeed^{-1}=\T/\tf$ is large
compared to the time scale of the system $\tsys$ describing the slowest relaxation
in the system. If the system is initially prepared in an arbitrary
distribution $p(x,0)$ it will relax into the stationary state at fixed
$\lambda_0$. This relaxation process occurs on a time scale that is much faster
than the time scale of the external driving. In the following we focus on the slow time scale on which the protocol is changing.
Therefore, we assume that the system has already relaxed into the stationary
state at $\lambda_0$. The density depending only on the slow time scale
\begin{equation}
  \label{eq:derivation_slow_driving_density}
  p(x,\tau;\pspeed) = \porder{0}(x,\tau) + \epss \porder{1}(x,\tau) + \order{\epss^2}
  \end{equation}
relaxes instantaneously into the stationary state
\begin{equation}
  \label{eq:derivation_slow_driving_density_zero_order}
  \porder{0}(x,\tau) = p^s(x,\lambda_\tau)
\end{equation}
at fixed $\lambda_\tau$, i.e., it fulfills
\begin{equation}
  \label{eq:derivation_slow_driving_FPE_zero_order}
  \FPOscaled \porder{0}(x, \tau) = 0.
\end{equation}
Equation~\eqref{eq:derivation_slow_driving_FPE_zero_order} follows by
inserting eq.~\eqref{eq:derivation_slow_driving_density}
into~\eqref{eq:derivation_scaled_FPE_new_notation} and by comparing the zeroth
orders in $\epss$. The time dependence of the
density~\eqref{eq:derivation_slow_driving_density_zero_order} is given through
the protocol $\lambda_\tau$. The density corresponds either to a NESS or an equilibrium state
at a fixed protocol $\lambda_\tau$.
If the density is that of a NESS at
fixed $\lambda_\tau$, the probability current
\begin{equation}
  \label{eq:derivation_slow_driving_probability_current}
  j(x,\tau;\pspeed) = \jorder{0}(x,\tau) + \epss \jorder{1}(x,\tau) + \order{\epss^2}
\end{equation}
converges to a finite value
\begin{equation}
  \label{eq:derivation_slow_driving_probability_current_zero_order_ss}
  \jorder{0}(x,\tau) =
  (\smobility \Ftot(x,\lambda_\tau) - \sD \partial_x) p^s(x,\lambda_\tau).
\end{equation}
In contrast if the density is an equilibrium
state at fixed $\lambda_\tau$, the driving is quasi-static and
the probability current vanishes such that $\jorder{0}(x,\tau)=0$ and
\begin{equation}
  \label{eq:slow_driving_probability_current_eq}
  j(x,\tau;\pspeed) = \epss \jorder{1}(x,\tau) + \order{\epss^2}.
\end{equation}

The time evolution operator~\eqref{eq:derivation_time_evolultion_operator}
converges to the leading order
\begin{equation}
  \label{eq:derivation_slow_driving_time_evolution_operator}
  \TEOorder{0}{x}{\tau_2}{\tau_1} \equiv \lim_{\epss\to 0} \TOexp{\int_{\tau_1}^{\tau_2}\dd{\tau}\epss^{-1}\FPOscaled},
\end{equation}
which satisfies
\begin{equation}
  \label{eq:derivation_slow_driving_time_evolution_operator_property}
  \TEOorder{0}{x}{\tau_2}{\tau_1} \rho(x,\tau_1) = p^s(x,\lambda_{\tau_2})
\end{equation}
for an arbitrary density
$\rho(x,\tau)$. Equation~\eqref{eq:derivation_slow_driving_time_evolution_operator_property}
shows that the time evolution operator
transforms any density into the stationary state at fixed $\lambda_\tau$. As a
consequence the leading order of the propagator is given by
\begin{equation}
  \label{eq:derivation_slow_driving_propagator}
  p(x_2,\tau_2\vert x_1,\tau_1) = p(x_2,\lambda_{\tau_2}) + \order{\epss}.
\end{equation}

We now use
eqs.~\eqref{eq:derivation_slow_driving_density},~\eqref{eq:derivation_slow_driving_probability_current}
and~\eqref{eq:derivation_slow_driving_propagator} to determine the leading
orders of the scaled mean values, eqs.~\eqref{eq:derivation_mean_value_instant_state_var}--\eqref{eq:derivation_mean_value_current_II},
their response
terms, their corresponding diffusion coefficients,
eqs.~\eqref{eq:derivation_diff_coeff_value_instant_state_var}--\eqref{eq:derivation_diffusion_coefficient_current_II}, as well as the
scaled total entropy production
rate~\eqref{eq:derivation_scaled_total_entropy_production_rate}.
We assume that all mean values and diffusion coefficients can be written
as
\begin{equation}
  \label{eq:derivation_fast_driving_general_scaling_mean}
  X(\T,\pspeed) \equiv \sum_{n=0}\left(\epss\right)^n X^{(n)}(\T,\pspeed)
\end{equation}
and
\begin{equation}
  \label{eq:derivation_fast_driving_general_scaling_diff_coeff}
  D_X(\T,\pspeed) \equiv \sum_{n=0}\left(\epss\right)^n D_X^{(n)}(\T,\pspeed)
\end{equation}
respectively, with mean values $X(\T,\pspeed)\in\{\Aim,\Acm,\Jtm{b,d}\}$ and
diffusion coefficients
$D_X(\T,\pspeed)\in\{\epss D_a(\T,\pspeed), D_A(\T,\pspeed), D_{J_{b,d}}(\T,\pspeed) \}$.
Their leading orders and the resulting quality factors are shown in
table~\ref{tab:ATDD_Slow_Driving}
(see~\ref{sec:appendix_limit_of_slow_driving} for details of the derivation).
\begin{table}[tbp]
  \centering
  \caption{Leading orders of the total entropy production rate, the response
    terms, the diffusion coefficents and the quality factors of the observables in the limit of slow
    driving. In this limit we have to distinguish whether a non-conservative force
    $f(\lambda_t)$ is applied to the system or not: if a non-conservative
    force is applied the system is in a NESS at fixed $\lambda_\tau$
    (NESS). However, if only a conservative force is applied, the system is in
    an equilibrium state for a fixed $\lambda_\tau$ (EQ).}
  \label{tab:ATDD_Slow_Driving}
  \begin{tabular}{llclllll}
    \toprule
    \multicolumn{1}{l}{Observable $X$} & \multicolumn{2}{l}{Response Term $\mathcal{R}_X$} &
\multicolumn{1}{l}{$D_X$} & \multicolumn{2}{l}{$\sigma(\T,\pspeed)$} &
                                                                       \multicolumn{2}{l}{$\mathcal{Q}_X$}\\
    \cmidrule(l){1-8}
          &\multicolumn{1}{c}{NESS}&\multicolumn{1}{c}{EQ} &&  \multicolumn{1}{l}{NESS} & \multicolumn{1}{l}{EQ} & \multicolumn{1}{l}{NESS} & \multicolumn{1}{l}{EQ}\\
    \midrule
    \midrule
    $X=\Aim$ & $\order{\epss^2}$ &$\order{\epss^2}$ & $\order{\epss^{-1}}$
                                       & $\order{1}$ & $\order{\epss^2}$
                                   &$\order{\epss^3}$ & $\order{\epss}$\\
    \midrule
    $X=\Acm$ & $\order{\epss^2}$ &$\order{\epss^2}$ &$\order{1}$ & $\order{1}$
                                                              &$\order{\epss^2}$
                                                        &$\order{\epss^2}$
                                                           & $\order{1}$\\
    \midrule
    $X=\Jtm{b}$ & $\order{\epss^4}$ & $\order{\epss^4}$&
                                                        $\order{\epss^2}$
                                       & $\order{1}$ &$\order{\epss^2}$ & $\order{\epss^2}$& $\order{1}$\\
    \midrule
    $X=\Jtm{d}$ & $\order{1}$ & $\order{\epss^4}$ &
                                                        $\order{1}$
                                       & $\order{1}$ &$\order{\epss^2}$
                                   &$\order{1}$ & $\order{\epss^2}$\\
    \bottomrule
  \end{tabular}
\end{table}
The response terms of the state variables vanish like $\epss^2$ due to
the fact that in the stationary state at fixed $\lambda_\tau$ the state variables
are invariant under a perturbation that scales the
time~\cite{dech18a,dech20a,koyu20}.
The same argument holds for the response term of the current $\Jtm{b}$, which
vanishes like $\epss^4$. The additional power of two comes from the
time-derivative of the increment. If a non-conservative force is applied, the
system is driven into a NESS, which implies that the response term of the current $\Jtm{d}$ is of
$\order{1}$ due to the
symmetry of the current under the scaling of time in a NESS. However, if only a
conservative force is applied, this symmetry does not longer
hold because the system is in an equilibrium state at fixed $\lambda_\tau$.
As a consequence the response term vanishes like $\epss^2$.
The diffusion coefficient of the state variable $\Acm$ and the current
$\Jtm{d}$ are of $\order{1}$ because their fluctuations are finite in the
stationary state at fixed $\lambda_\tau$. The instantaneous state variable diverges
proportional to $\epss^{-1}$ due to the factor of $\T$ in the
definition of its diffusion constant. The diffusion coefficient of the
current $\Jtm{b}$ vanishes like $\epss$ due to the time-derivative of
its increment. The total entropy production rate is of $\order{1}$, if a
non-conservative force is applied. In this case the probability currents do not vanish and are of
$\order{1}$. In contrast if the system is only driven by a conservative force,
the total entropy production rate is of
$\order{\epss^2}$. In this case the system is in an equilibrium state at fixed
$\lambda_\tau$ and hence, the probability currents vanish like $\epss$.
Combining these results yields to the leading orders of the quality factors:
if a non-conservative force is applied, all quality factors except
the quality factor for the current $\Jtm{d}$ vanish. In contrast, if only a
conservative force is applied, the quality
factors of the state variable $\Acm$ and of the current $\Jtm{b}$ are of
$\order{1}$. The other quality factors vanish asymptotically.
To summarize, a useful estimate for the entropy production rate is only possible for the
current $\Jtm{d}$, if a non-conservative force is applied or for
both, the state variable $\Acm$ and the current $\Jtm{b}$, if the system is
driven by a conservative force only.

\section{Systems with discrete states: three-state model}
Our main results, the scaling of the quality factors in
table~\ref{tab:ATDD_Fast_Driving} for fast driving and in
table~\ref{tab:ATDD_Slow_Driving} for slow driving, hold generically not only
for overdamped Langevin systems but also for systems with a set of discrete
states described by a Markovian dynamics as the derivation of the scaling of
the quality factors follows the same steps presented in
sections~\ref{sec:fast_driving} and~\ref{sec:slow_driving}. The
dynamics for the probability to find the system in a discrete state $i$ is
described by the master equation
\begin{equation}
  \label{eq:illustration_three_level_system_master_equation}
  \partial_t p_i(t;\pspeed) = -\sum_j j_{ij}(t;\pspeed)
\end{equation}
with probability current
\begin{equation}
  \label{eq:illustration_three_level_system_master_equation_current}
  j_{ij}(t;\pspeed) \equiv p_i(t;\pspeed)k_{ij}(\lambda_t) - p_j(t;\pspeed)k_{ji}(\lambda_t),
\end{equation}
where we introduced the dependence with respect to the speed parameter $\pspeed$ as the
second argument for both, the probability to find the system in a state $i$
and the probability current between two states $i$ and $j$.
The transition rates $k_{ij}(\lambda_t)$ between two states $i$ and $j$ are
time-dependent through the protocol $\lambda_t$ and fulfill the local
  detailed balance condition
\begin{equation}
  \label{eq:illustration_three_level_system_local_detailed_balance_condition}
  \frac{k_{ij}(\lambda_t)}{k_{ji}(\lambda_t)} = \exp{-\beta
  \left[E_j(\lambda_t) - E_i(\lambda_t)\right] - \mathcal{A}_{ij}(\lambda_t)},
\end{equation}
where $\beta$ denotes the inverse temperature, $E_i(\lambda_t)$ denotes the time-dependent energy of state $i$ and
$\mathcal{A}_{ij}(\lambda_t)$ is a driving affinity, which drives the system additionally
to the time-dependent energies into a non-equilibrium state.

\begin{figure}[tbp]
  \centering
  \includegraphics[width=\textwidth]{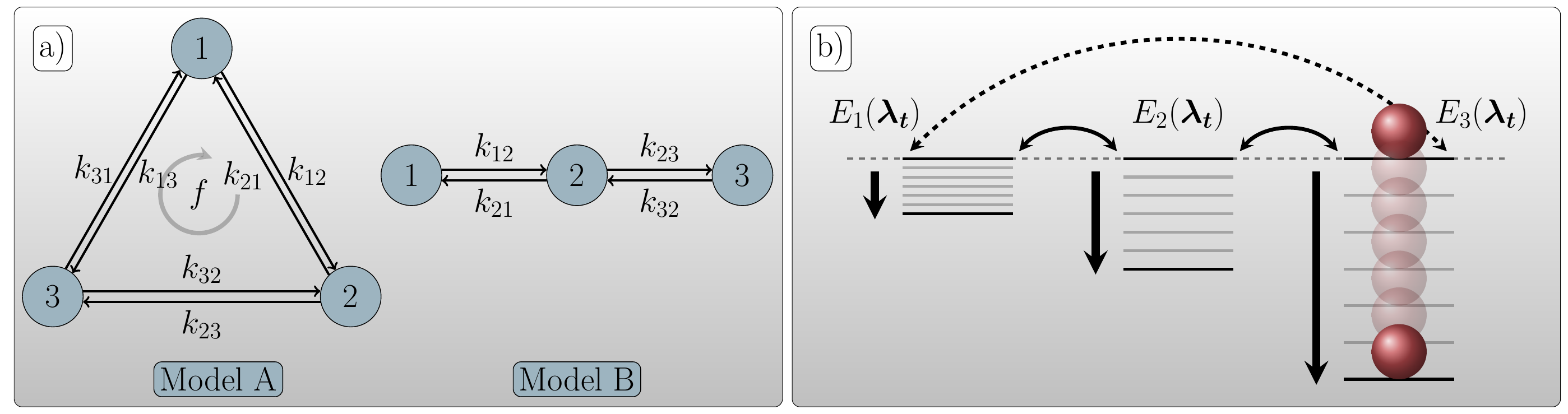}
  \caption{Topology of the two models A and B for the three-state system (a) and
    schematic of the three-state system with time-dependent energy levels
    (b). In model A there is a link between states 1 and 3 such that a NESS can be
    reached by applying a non-conservative force $f$. In model B there is no
    link betweenstates 1 and 3, which limits the net number of transitions up to a finite
    observation time $\T$ between two
    states. The three energy levels have initially the same value $E_i(0)=0$
    and are decreased over time to a fixed final value $E_i(\lambda_{\tau_f})=-E^0_i
    \tau_f^2$.}
  \label{fig:illustration_three_level_system_model_schematic}
\end{figure}

As an example we consider a system with three discrete states, where
the energy levels of the states are driven time-dependently through a protocol
$\lambda_t$. The
topology of this network is shown in
Fig.~\ref{fig:illustration_three_level_system_model_schematic}a).
We distinguish between two models: model A contains a link between state 1
and 3. In addition to the time-dependent driving of the energy levels, the
system is driven by a constant non-conservative force $f$. In model B there is no
link between state 1 and 3. As a consequence the net number of transitions
between two states is zero or $\pm 1$, which implies that their fluctuations are not
time-extensive.
The energy levels of the three states
\begin{equation}
  \label{eq:illustration_three_level_system_definition_energy_levels}
  E_i(\lambda_t) \equiv - E^{0}_i \lambda_t,
\end{equation}
are driven by a quadratic protocol
\begin{equation}
  \label{eq:illustration_three_level_system_definition_protocol}
  \lambda_t \equiv \left(\pspeed t\right)^2,
\end{equation}
where $E^{0}_i$ is the amplitude of the driving and $\pspeed$ is the speed
parameter. The rates are chosen according to the local detailed
  balance condition~\eqref{eq:illustration_three_level_system_local_detailed_balance_condition} and read
\begin{align}
  \label{eq:illustration_three_level_system_definition_rates_ij}
  k_{ij}(\lambda_t) &\equiv k_0^{ij}\exp(-\beta \left[E_j(\lambda_t) -
                      E_j(\lambda_t)\right]/2 - f/6),\\
  \label{eq:illustration_three_level_system_definition_rates_ji}
  k_{ji}(\lambda_t) &\equiv k_0^{ij}\exp(\beta \left[E_j(\lambda_t) -
  E_j(\lambda_t)\right]/2 + f/6),
\end{align}
where we have chosen the driving affinity as a constant
$\mathcal{A}_{ij}(\lambda_t)=-\mathcal{A}_{ji}(\lambda_t)=f/3$.
The rate amplitudes $k_0^{ij}$ determine time scale of the system $\tsys$. In the following, we
set all the rate amplitudes to the same value $k_0^{ij}\equiv k_0 =
1$ and choose all other parameters $\beta$, $E_i(\lambda_t)$ and $f$ of $\order{1}$.  As a consequence all relaxation
times in system are of the same order of magnitude and hence, we are able to
choose $\tsys \equiv 1/k_0 = 1$ as outlined in
section~\ref{sec:time_scale_separation}. Moreover, we choose the initial distribution
$p_i(0)$ as the stationary state at fixed $\lambda_0$ at the beginning of the driving.

In the following, we analyze the quality factors for both models, A and B in
the limits of fast and slow driving for several types of observables.
As an example for the instantaneous state variable, we consider the variable
\begin{equation}
  \label{eq:illustration_three_level_system_instantaneous_state_variable_example_probability}
  \Aif^{i} = \delta_{i,i(\T)}.
\end{equation}
Its mean value is the probability to find the system in state $i$ at the end
of the observation time $\T$. Here, $\delta_{i,i(t)}$ is one, if the trajectory $i(t)$ is in state $i$ and
zero, otherwise. We further analyze the time-average over this variable
\begin{equation}
  \label{eq:illustration_three_level_system_time_averaged_state_variable_example_fraction_of_time}
  \Acf^{i} = \frac{1}{\T}\int_0^\T\dd{t}\delta_{i,i(t)},
\end{equation}
which is the overall fraction of time the system has spent in state $i$ up to
the finite observation time $\T$. For the current-observables we analyze the
power
\begin{equation}
  \label{eq:illustration_three_level_system_residence_current_example_power}
  P^i_\T \equiv \frac{1}{\T}\int_0^\T\dd{t} \dot{E}_i(\lambda_t)\delta_{i,i(t)}
\end{equation}
exerted at energy level $i$, which is an example for the current $\Jtf{b}$ and
the rate of directed number of transitions between state $i$ and $j$
\begin{equation}
  \label{eq:illustration_three_level_system_velocity_current_example_number_of_transitions}
  J^{ij}_\T \equiv \frac{1}{\T}\int_0^\T\dd{t} \left[\dot{n}_{ij}(t) - \dot{n}_{ji}(t)\right],
\end{equation}
which is an example for the current $\Jtf{d}$. Here, $n_{ij}(t)$ denotes the
number of transitions between states $i$ and $j$ up to time $t$ along a
trajectory $i(t)$. The average value of
eq.~\eqref{eq:illustration_three_level_system_velocity_current_example_number_of_transitions}
is the time-averaged probability current between state $i$ and $j$.

In the following we plot \textit{inter alia} $\mathcal{Q}_X/\epsilon_{s,f}^n$ against
$\epsilon_{s,f}^{-1}$ to analyze the scaling of the quality factor for
an observable $X$. Here, $\mathcal{Q}_X/\epsilon_{s,f}^n$ converges to a constant value
for the correct power $n$ (see table~\ref{tab:ATDD_Fast_Driving}
and~\ref{tab:ATDD_Slow_Driving}) in the limit of fast-driving
$\epsf^{-1}\to\infty$ and slow-driving $\epss^{-1}\to\infty$, i.e., $\epsf\to
0$ and $\epss\to 0$, respectively.

\subsection{Model A}
The topology of model A in fig.~\ref{fig:illustration_three_level_system_model_schematic}a) allows the system to reach a NESS by applying
a non-conservative force $f$ or to converge to an equilibrium system by 
applying only a conservative force. These distinct two
cases are especially relevant for the limit of slow driving.

\begin{figure}[tbp]
  \centering
  \includegraphics[width=\textwidth]{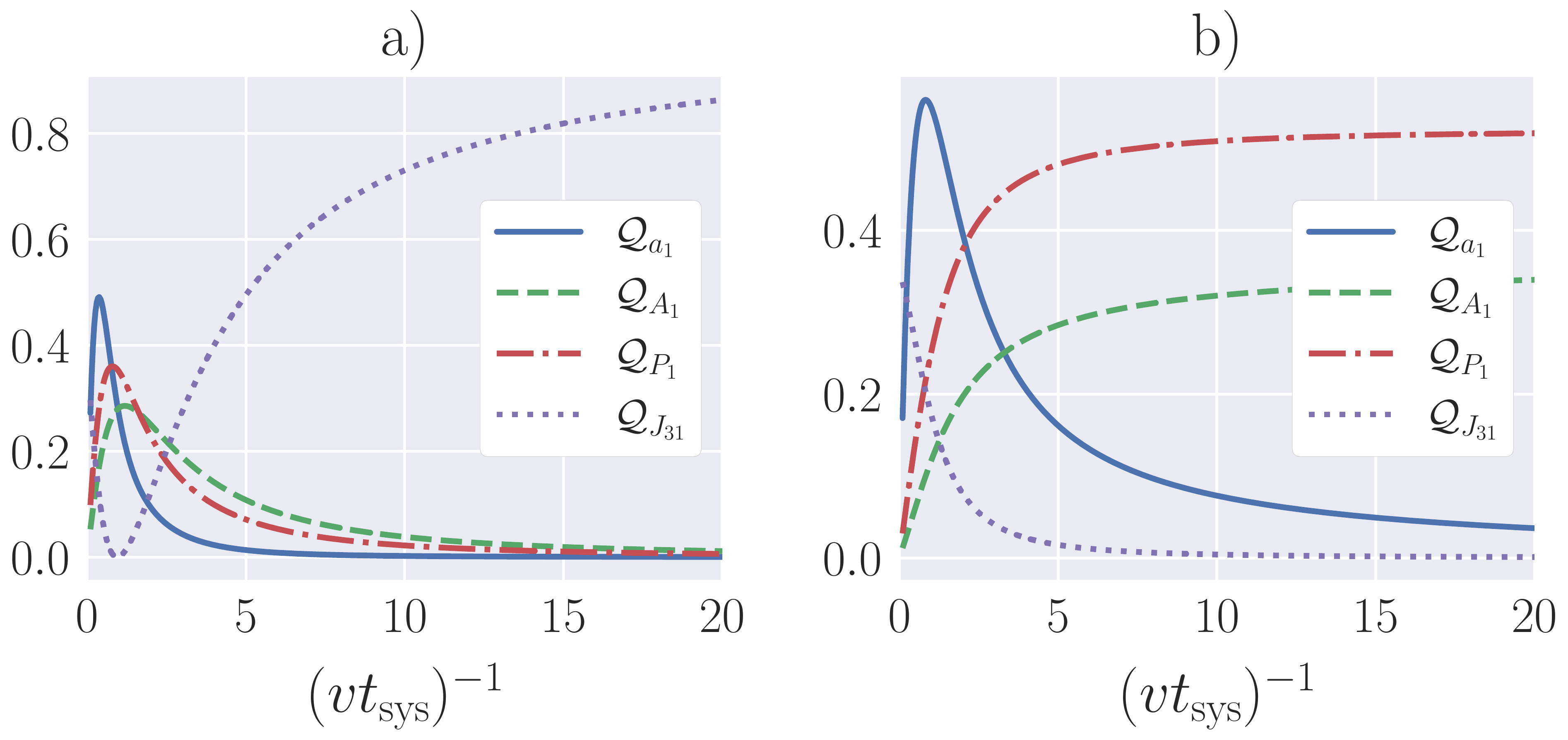}
  \caption{Quality factors of the probability $a_1(\T,\pspeed)$ to find the system in state 1,
    of the fraction of time $A_1(\T,\pspeed)$ the system has spent in state 1,
    of the power $P_1(\T,\pspeed)$
    applied to state 1 and of the current
    $J_{31}(\T,\pspeed)$ between states 3
    and 1. The quality
    factors are plotted against $(\pspeed\tsys)^{-1}$ for $f=1.5$ (a) and
    for $f=0$ (b). Here, we have set $\beta=1.0$, $E^0_1=0.5$, $E^0_2=1.0$ and $E^0_3=2$.}
  \label{fig:illustration_three_level_system_model_A_all_T}
\end{figure}

Figure~\ref{fig:illustration_three_level_system_model_A_all_T}
shows the quality factors of the different types of observables defined in
eqs.~\eqref{eq:illustration_three_level_system_instantaneous_state_variable_example_probability}--\eqref{eq:illustration_three_level_system_velocity_current_example_number_of_transitions}
for a finite non-conservative force $f\ge 0$ (a) and for a vanishing
non-conservative force $f=0$ (b).
Comparing the results for fast driving in table~\ref{tab:ATDD_Fast_Driving} and
for slow driving in table~\ref{tab:ATDD_Slow_Driving} for the three-state model let us
conclude that either the current $J_{31}(\T,\pspeed)\equiv\expval{J_\T^{31}}$
between state 3 and 1 or the
time-averaged state variable $A_1(\T,\pspeed)\equiv\expval{\Acf^{1}}$ as well
as the power $P_1(\T,\pspeed)\equiv\expval{P^{1}_\T}$ are the best choice to infer the total entropy production in the
respective limiting cases. However, the instantaneous state variable, the
probability $a_1(\T,\pspeed)\equiv\expval{\Aif^{1}}$ to find the system in state 1, is not
an optimal choice for both limiting cases as its quality factor vanishes as
shown in fig.~\ref{fig:illustration_three_level_system_model_A_all_T}a) and b). 
In contrast, we expect that the quality factor for the
instantaneous state variable has a maximum and is of
$\order{1}$ for a speed of driving comparable with the time scale
of the system, i.e., $(\pspeed\tsys)^{-1}\sim 1$. This can be seen for the three-state model in
fig.~\ref{fig:illustration_three_level_system_model_A_all_T}a) and b), where
the instantaneous state variable yields about $50\%$ of the total entropy
production rate.

\begin{figure}[tbp]
  \centering
  \includegraphics[width=\textwidth]{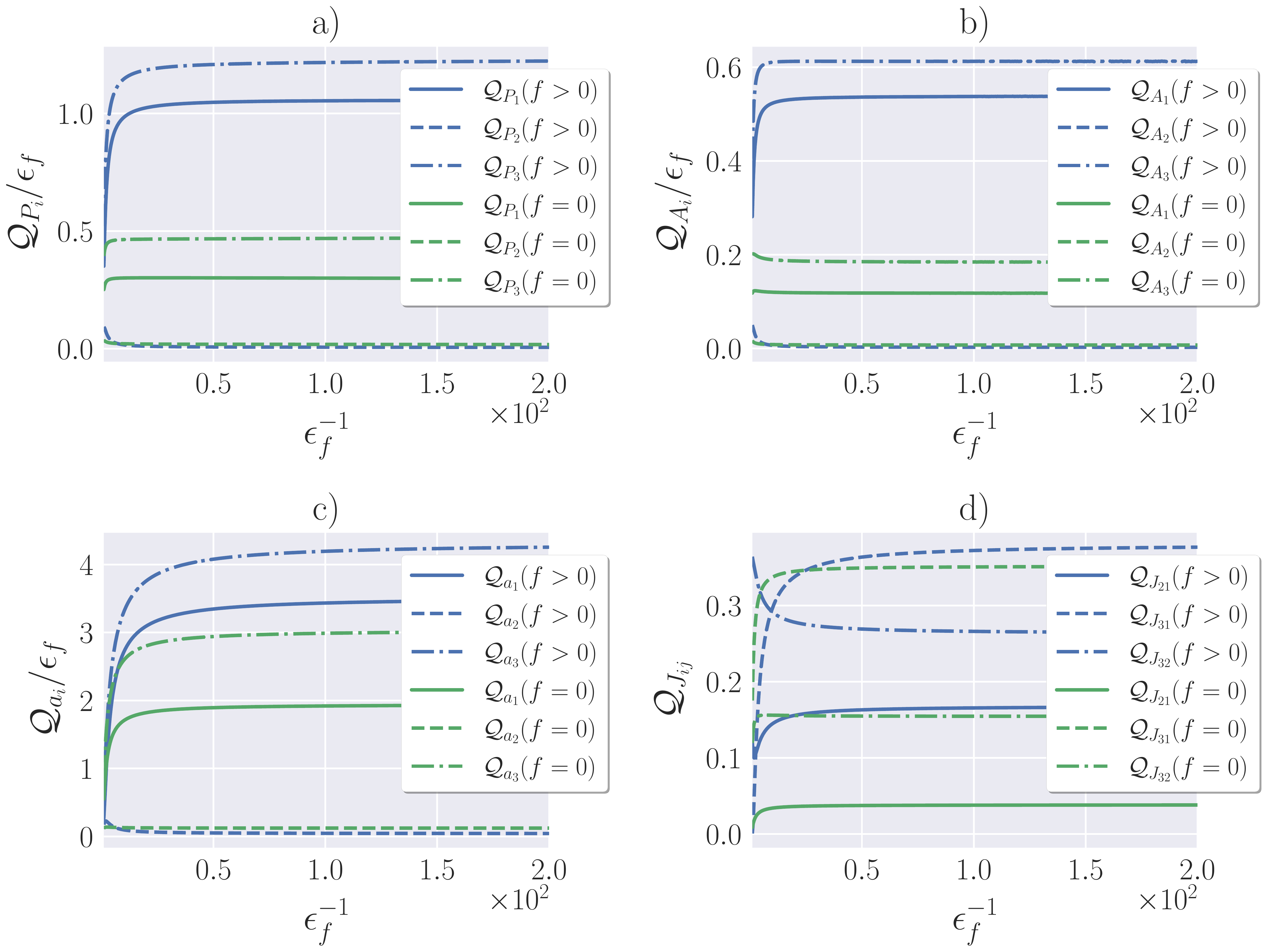}
  \caption{Quality factors in the limit of fast driving for the power (a), for the fraction of time spent
    in a certain state (b), for the probability to find the system in a
    certain state (c) and for the time-averaged probability current between two
  states (d). The quality factors are plotted against the parameter
  $\epsf^{-1}$ and shown for a finite non-conservative force $f=1.5>0$ and for
a vanishing force $f=0$. Here, we have set
    $\beta=1.0$, $E^0_1=0.5$, $E^0_2=1.0$ and $E^0_3=2$.}
  \label{fig:illustration_three_level_system_model_A_fast_driving}
\end{figure}

Next, we analyze the quality factors for the observables defined in
eqs.~\eqref{eq:illustration_three_level_system_instantaneous_state_variable_example_probability}--\eqref{eq:illustration_three_level_system_velocity_current_example_number_of_transitions}
in the limit of fast-driving. The quality factors for the power, for the
fraction of time the system has spent in a certain state, for the probability
to find the system in a state and for the time-averaged current between two
states are shown in fig.~\ref{fig:illustration_three_level_system_model_A_fast_driving}a)--d).
As predicted by table~\ref{tab:ATDD_Fast_Driving}, the quality factors for the
state variables $A_i(\T,\pspeed)$ and $a_i(\T,\pspeed)$ and the current depending on the residence
time $P_i(\T,\pspeed)$ are proportional to $\epsf$ as
their quality factors divided by $\epsf$ converge to a constant value in the
limit $\epsf^{-1}\to\infty$ (see
fig.~\ref{fig:illustration_three_level_system_model_A_fast_driving}a)--c)). The
quality factor for the current $J_{ij}(\T,\pspeed)$
converges to a constant value and is of $\order{1}$ as shown in
fig~\ref{fig:illustration_three_level_system_model_A_fast_driving}d). The
scaling of these quality factors are independent of the force $f$.

\begin{figure}[tbp]
  \centering
  \includegraphics[width=\textwidth]{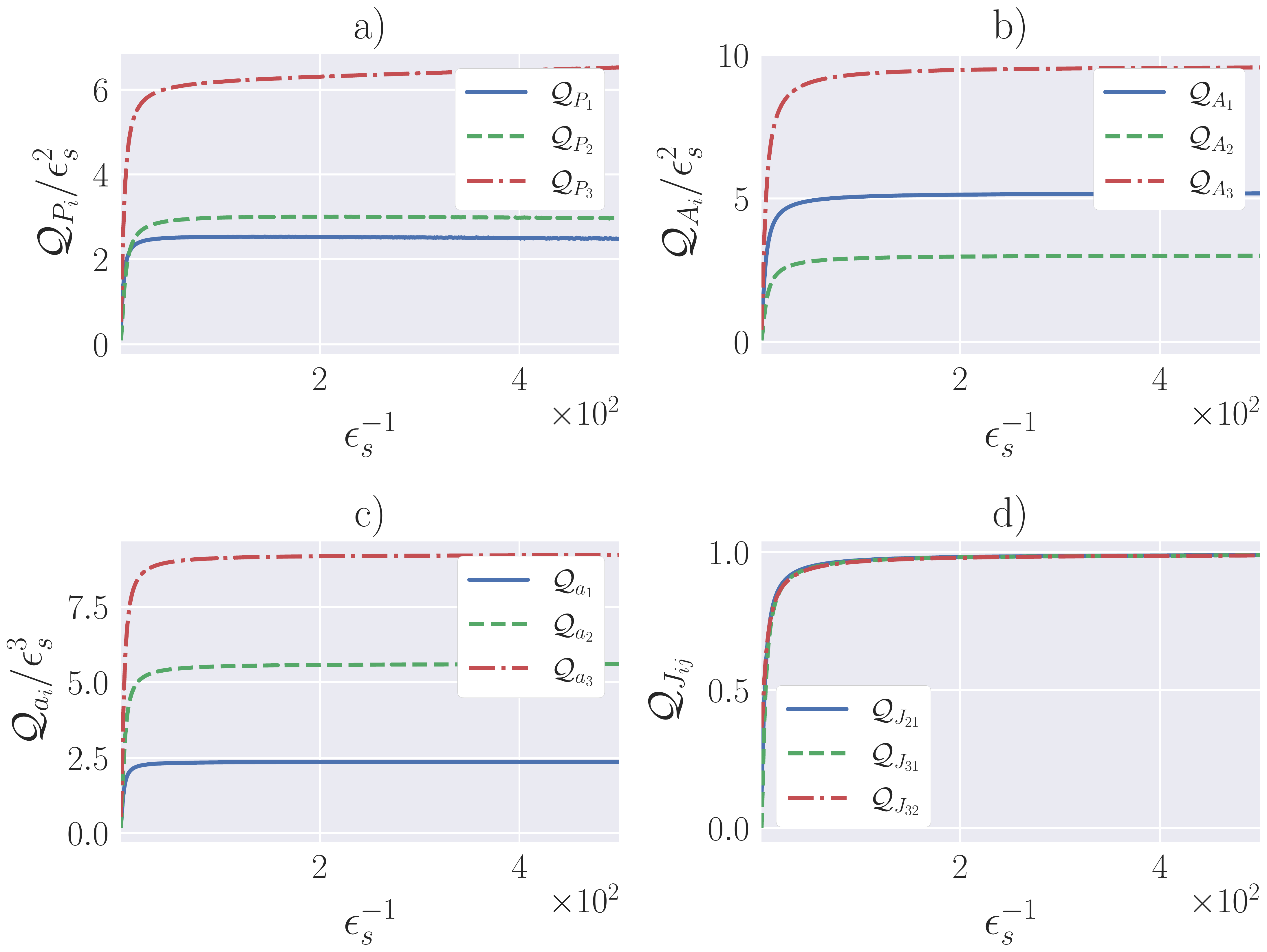}
  \caption{Quality factors in the limit of slow driving for the power (a), for the fraction of time spent
    in a certain state (b), for the probability to find the system in a
    certain state (c) and for the time-averaged probability current between two
  states (d). The quality factors are plotted against the parameter
  $\epss^{-1}$ and shown for a finite force $f=1.5>0$. Here, we have set
    $\beta=1.0$, $E^0_1=0.5$, $E^0_2=1.0$ and $E^0_3=2$.}
  \label{fig:illustration_three_level_system_model_A_slow_driving_f_neq_0}
\end{figure}

In contrast, in the limit of slow driving the scaling of the quality
factors depend on the non-conservative force $f$. For $f>0$, the system converges to a
NESS at a constant $\lambda_t$, whereas for a vanishing force $f=0$ the system
converges to an equilibrium state at constant $\lambda_t$. We first focus on
the case of a non-vanishing force $f>0$. The quality factors for the power, for the
fraction of time the system has spent in a certain state, for the probability
to find the system in a state and for the time-averaged current between two
states are shown in fig.~\ref{fig:illustration_three_level_system_model_A_slow_driving_f_neq_0}a)--d).
The quality factors for the current $P_i(\T,\pspeed)$ and the time-averaged state variable
$A_i(\T,\pspeed)$ scale like $\epss^2$, whereas it scales for the instantaneous state
variable like $\epss^3$. In contrast, the only the quality factor of the
current $J_{ij}(\T,\pspeed)$ yields to a quality factor of $\order{1}$. There
are three quality factors of the time-averaged probability current for each link:
$\mathcal{Q}_{J_{21}}$, $\mathcal{Q}_{J_{31}}$ and
$\mathcal{Q}_{J_{31}}$. While all three of the quality factors are different in the
region of small $\epss^{-1}$, where the slow-driving limit is not yet reached,
they converge asymptotically identical to the same value in the limit of slow
driving ($\epss^{-1}\to\infty$). This can be understood as follows. When the
system is driven slowly enough, it passes different NESSs in the course of time. In
each NESS all three currents are identical. As a
consequence the quality factors must also be identical. To summarize, the optimal observable leading to a useful estimate for the total
entropy production rate is a current $\Jtm{d}$ depending on the velocity or,
equivalently, depending on the number of transitions between two discrete
states. All other observables yield a quality factor that vanishes at least of
order $\epss^2$ as predicted in table~\ref{tab:ATDD_Slow_Driving}.

\begin{figure}[tbp]
  \centering
  \includegraphics[width=\textwidth]{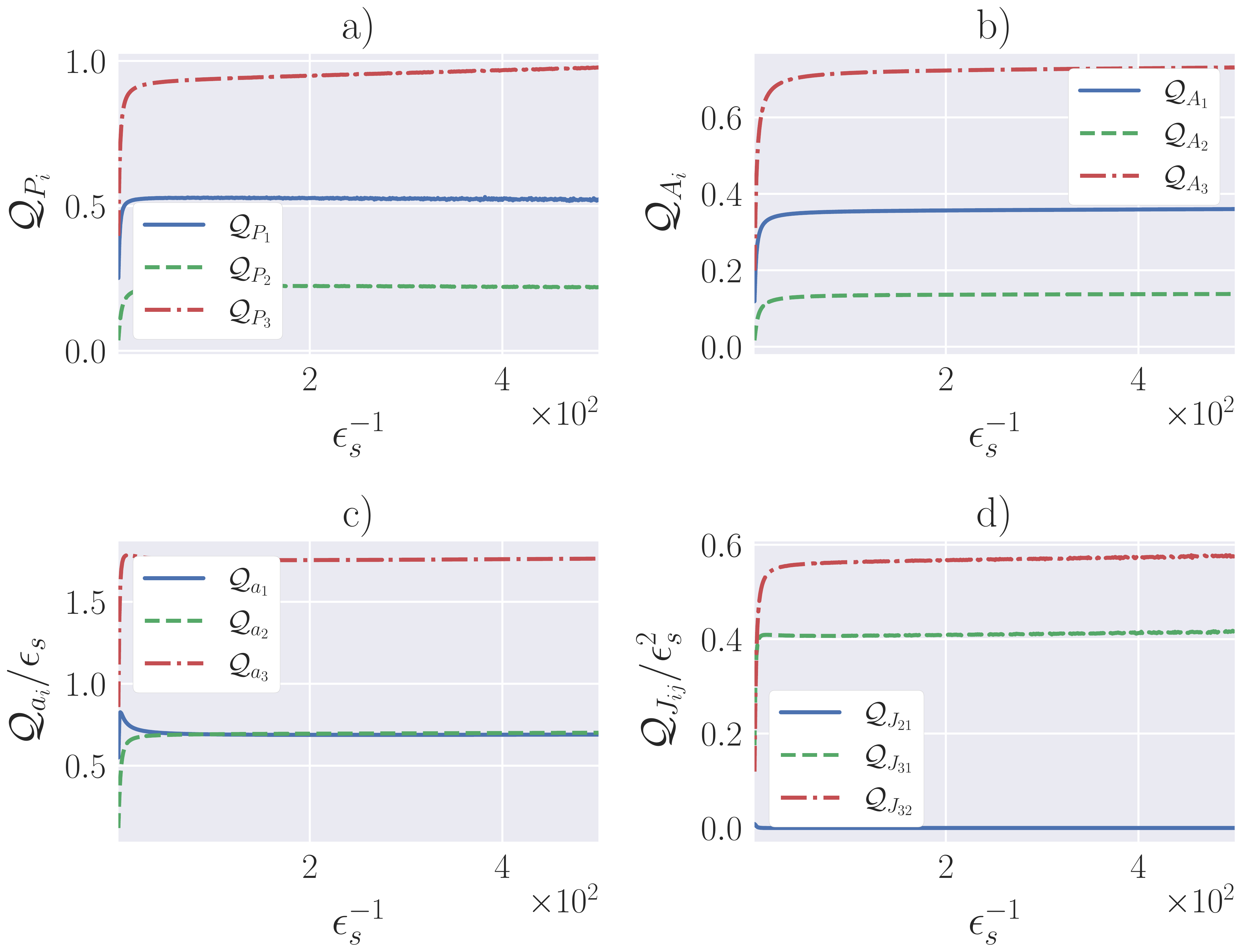}
  \caption{Quality factors in the limit of slow driving for the power (a), for the fraction of time spent
    in a certain state (b), for the probability to find the system in a
    certain state (c) and for the time-averaged probability current between two
    states (d). The quality factors are plotted against the parameter
    $\epss^{-1}$ and shown for a finite force $f=0$. Here, we have set
    $\beta=1.0$, $E^0_1=0.5$, $E^0_2=1.0$ and $E^0_3=2$.}
  \label{fig:illustration_three_level_system_model_A_slow_driving_f_eq_0}
\end{figure}

Next, we consider the limit of slow driving for a vanishing force $f=0$, where
the system is in an equilibrium state at fixed $\lambda_t$. The quality factors for the power, for the
fraction of time the system has spent in a certain state, for the probability
to find the system in a state and for the time-averaged current between two
states are shown in fig.~\ref{fig:illustration_three_level_system_model_A_slow_driving_f_eq_0}a)--d).
In contrast to the case with $f>0$, where the quality factors of the current $P_i(\T,\pspeed)$ and
the state variable $A_i(\T,\pspeed)$ vanish like $\epss^2$ (see
fig.\ref{fig:illustration_three_level_system_model_A_slow_driving_f_neq_0}a)
and b)),
they both converge to a value of $\order{1}$  as
shown in
fig.~\ref{fig:illustration_three_level_system_model_A_slow_driving_f_eq_0}a)
and b). In the latter case, the time-average state
variable $A_3(\T,\pspeed)$ yields over $60\%$ of the total entropy
production rate. The quality factor for the power $P_3(\T,\pspeed)$ even nearly
saturates due to the fact that the total power $P_\mathrm{tot}(\T,\pspeed)
\equiv \sum_i P_i(\T,\pspeed)$ converges to the entropy production rate in the
limit of slow driving. The power $P_3(\T,\pspeed)$ contributes the most to the
total power (due to $E^0_3 > E^0_2 > E^0_1$ as sketched in fig.~\ref{fig:illustration_three_level_system_model_schematic}b)) and hence, yields to the best estimate for the total entropy
production rate. As shown in fig.~\ref{fig:illustration_three_level_system_model_A_slow_driving_f_eq_0}c)
and d) the quality factors for the state variable $a_i(\T,\pspeed)$ and the
for the current $J_{ij}(\T,\pspeed)$ vanish like $\epss$ and $\epss^2$,
respectively. This is an important contrast to the case $f>0$, where the
quality factor for the current $J_{ij}(\T,\pspeed)$ is of $\order{1}$.

\subsection{Model B}
\begin{figure}[tbp]
  \centering
  \includegraphics[width=\textwidth]{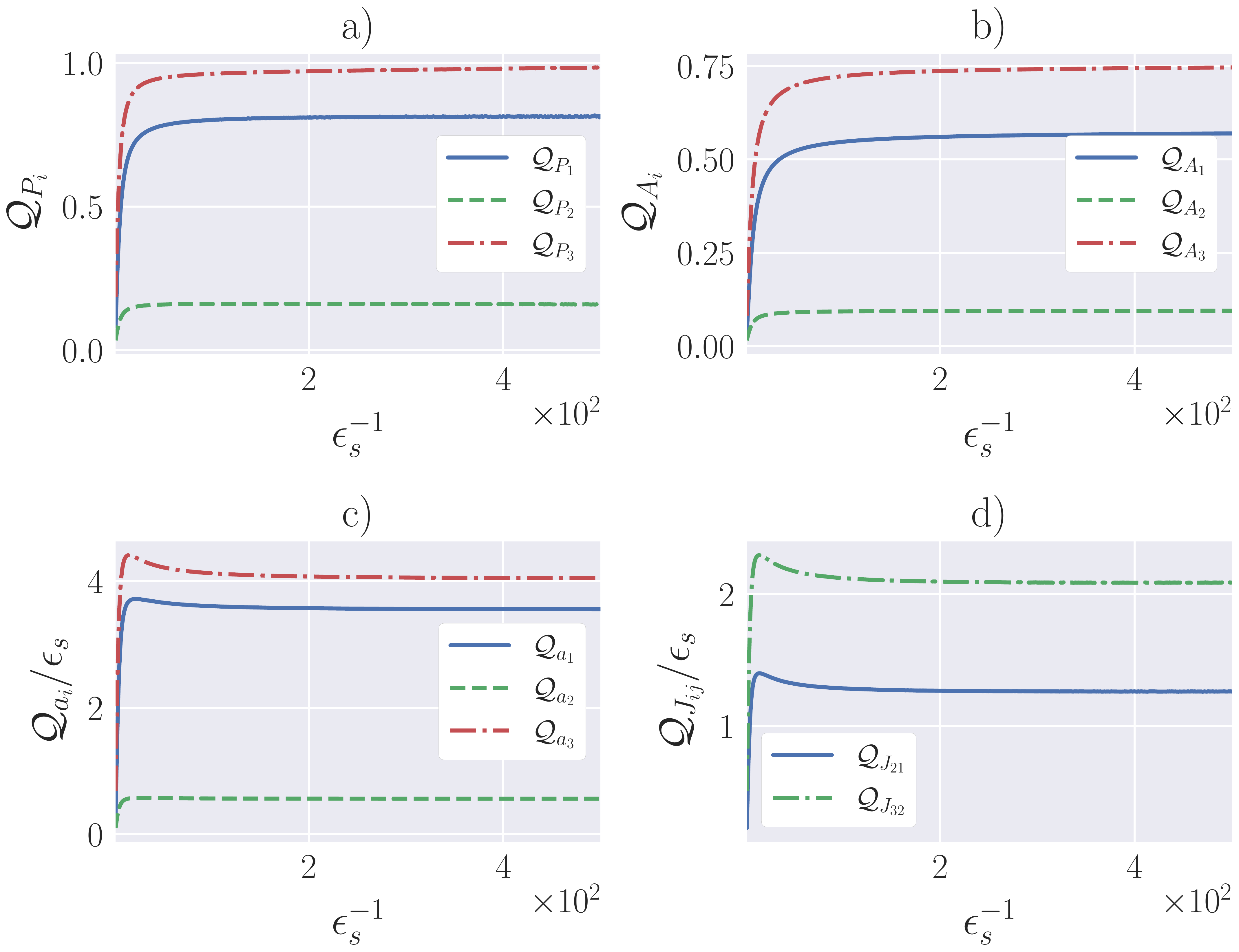}
  \caption{Quality factors in the limit of slow driving for the power (a), for the fraction of time spent
    in a certain state (b), for the probability to find the system in a
    certain state (c) and for the time-averaged probability current between two
    states (d). The quality factors are plotted against the parameter
    $\epss^{-1}$. Here, we have set
    $\beta=1.0$, $E^0_1=0.5$, $E^0_2=1.0$ and $E^0_3=2$.}
  \label{fig:illustration_three_level_system_model_B_slow_driving}
\end{figure}

Model B cannot be driven into a NESS due to its topology depicted in
fig.~\ref{fig:illustration_three_level_system_model_schematic}a). Hence, the system
can only reach an equilibrium state, which leads to the generic scaling for
the quality factors in the limit of slow driving as shown in
table~\ref{tab:ATDD_Slow_Driving} (second columns).
However, in this system the topology of the network leads to deviations of the
predicted scaling
behavior of these quality factors, which are shown in fig.~\ref{fig:illustration_three_level_system_model_B_slow_driving}a)--d).
Only the state variable $A_i(\T,\pspeed)$ and the current
$P_i(\T,\pspeed)$ yield an estimate for the entropy production of
$\order{1}$ as shown in
fig.~\ref{fig:illustration_three_level_system_model_B_slow_driving}a) and
b). The quality factors for the instantaneous variable $a_i(\T,\pspeed)$ and
for the current $J_{ij}(\T,\pspeed)$ vanish. However, the quality factor of
the current $J_{ij}(\T,\pspeed)$ does not vanish like $\epss^2$ as generically
predicted in table~\ref{tab:ATDD_Slow_Driving} but scales like $\epss$. This
circumstance follows from the fact that the net number of
transitions between two states and consequently also their fluctuations
cannot become arbitrary large. As a consequence, the diffusion coefficient of the current is not
of $\order{1}$ but vanishes proportional to $\epss$. This leads to the
modified scaling of $\order{\epss}$ as shown in fig~\ref{fig:illustration_three_level_system_model_B_slow_driving}d).

\section{Conclusion}
In this paper, we have analyzed the quality of the thermodynamic uncertainty
relation for the limiting cases of fast driving and slow driving. In the limit
of fast driving, the generic optimal observable is the current-observable
depending on the velocity. The quality factors of all other observables vanish
asymptotically.
We have further shown that in the limit of fast driving a current proportional
to the total entropy production rate can saturate the
uncertainty relation. In the limit of slow driving, one has to distinguish whether a driving
affinity is additionally applied to the system or not in order to choose the optimal
observable. If the system is driven by a driving affinity the optimal
observable is the current depending on the velocity. However, if there is no
driving affinity only the current depending on the residence time or the
time-averaged state variable yields an useful estimate for the total entropy
production rate. All other quality factors vanish generically with a power law
in the ratio of the relevant time scales. The quality factor of the instantaneous state variable
vanishes in both limiting cases. However, as we have illustrated for a three-level system it still can yield a useful estimate when the speed of driving is
comparable with the relaxation time scales in the system. Last but not least, an
analysis of the quality factors in the three-state model shows that depending on the topology of
the system deviations of the generic scaling of the quality factors can occur. 

With these results we have introduced first steps for optimizing
inference schemes using the thermodynamic uncertainty relation for
time-dependent driving. We have focused on the scaling behavior of various
classes of observables. In a next step, one could investigate which observable
is optimal within each class. Moreover, it would be possible to use
a superposition of two observables or to
involve correlations between them in order to optimize the bounds on entropy
production~\cite{dech21}. Lastly, a further open question is how the quality
of the TUR behaves as a function of system size in more complex models.

\begin{appendix}

  \section{Diffusion Coefficients and Correlation Functions}
  \subsection{Diffusion Coefficients}
  \label{sec:appendix_diffusion_coefficients}
  In this section, we give the explicit expressions for the diffusion
  coefficients of the observables defined in eqs.~\eqref{eq:derivation_instant_state_var}--\eqref{eq:derivation_current_II}.
  The diffusion coefficients of the state variables~\eqref{eq:derivation_instant_state_var}
  and~\eqref{eq:derivation_time_averaged_state_var} in terms of the scaled time
  $\tau$ are given by
  \begin{equation}
    \label{eq:derivation_diff_coeff_value_instant_state_var}
    D_a(\T,\pspeed) = \T\left(\int\dd{x} a^2(x,\lambda_{\tf}) p(x,\tf;\pspeed) - \Aim^2\right)/2
  \end{equation}
  and
  \begin{align}
    \label{eq:derivation_diff_coeff_value_time_averaged_state_var}
    D_A(\T,\pspeed) =&
    \frac{\T}{\tf^2}\int_0^{\tf}\dd{\tau}\int_0^{\tau}\dd{\tau'}\int\dd{x}\int\dd{x'}a(x,\lambda_{\tau})p(x,\tau\vert
    x',\tau')a(x',\lambda_{\tau})p(x',\tau;\pspeed)\nonumber\\
    &- \T/2 \Acm^2,
  \end{align}
  respectively. Analogously, the diffusion coefficient of the the
  current~\eqref{eq:derivation_current_I} is given by
  \begin{align}
    \label{eq:derivation_diffusion_coefficient_current_I}
    D_{J^b}(\T,\pspeed)  =&
    \frac{\pspeed}{\tf}\int_0^{\tf}\dd{\tau}\int_0^{\tau}\dd{\tau'}\int\dd{x}\int\dd{x'}\dot{b}(x,\lambda_{\tau})p(x,\tau\vert
    x',\tau')\dot{b}(x',\lambda_{\tau})p(x',\tau;\pspeed)\nonumber\\
    &- \T/2 \Jtm{b}^2.
  \end{align}
  In order to write the diffusion coefficient of the current in eq.~\eqref{eq:derivation_current_II} in terms
  of correlation functions between state functions, we use the
  relation~\eqref{eq:derivation_relation_x_dot} in~\ref{sec:appendix_correlation_functions}.
  Plugging eq.~\eqref{eq:derivation_relation_x_dot} into the diffusion
  coefficient~\eqref{eq:derivation_diffusion_coefficient} of the current
  depending on the velocity and changing to the time scale $\tau$ yields
  \begin{align}
    \label{eq:derivation_diffusion_coefficient_current_II}
    D_{J^d}(\T,\pspeed) = &\frac{1}{\tsys\tf}\int_0^{\tf}\dd{\tau}\int\dd{x} \sD
                            d^2(x,\lambda_\tau)p(x,\tau;\pspeed) \nonumber \\
                          &+
                            \frac{(\pspeed\tsys)^{-1}}{\tsys\tf}\int_0^{\tf}\dd{\tau}\int_0^\tau\dd{\tau'}\int\dd{x}\int\dd{x'}\scaledsymbol{J}(x,\tau)p(x,\tau\vert
                            x', \tau')  \nonumber \\
    &\hspace{3cm}\times\{ \scaledsymbol{J}(x',\tau') p(x',\tau';\pspeed) - 2\sD
    \left[d(x',\lambda_{\tau'}) p(x',\tau';\pspeed)\right]' \}\nonumber\\
                          & - \frac{\T}{2} \Jtm{d}^2
  \end{align}
  with
  \begin{equation}
    \label{eq:derivation_scaled_J_t}
    \scaledsymbol{J}(x,\tau) \equiv d(x,\lambda_\tau) j(x,\tau;\pspeed)/p(x,\tau;\pspeed) + \sD \{d(x,\lambda_\tau)p(x,\tau;\pspeed)\}'/p(x,\tau;\pspeed).
  \end{equation}

  \subsection{Correlation Functions}
  \label{sec:appendix_correlation_functions}
  Throughout this section, we use the original notation for the Fokker-Planck equation
  introduced in eq.~\eqref{eq:derivation:Fokker-Planck-Equation} and not the
  time-scaled notation introduced in section~\ref{sec:time_scale_separation}.
  In the following, we derive the relation
  \begin{align}
    \label{eq:derivation_relation_x_dot}
    \expval{d(x_t,\lambda_t)\circ\dot{x}_t
    d(x_{t'},\lambda_{t'})\circ\dot{x}_{t'}} &=\,  2D \expval{d^2(x_t,\lambda_t)}\delta(t-t') \nonumber \\
                                                                            &+ \theta(t-t') \expval{J(x_{t'},t')\left[J(x_t,t) -
                                                                              2D \{d(x_t,\lambda_t)p(x_t,t;\pspeed)\}'/p(x_t,t)\right]}
                                                                              \nonumber \\
                                                                            &+ \theta(t'-t) \expval{J(x_t,t)\left[J(x_{t'},t') -
                                                                              2D \{d(x_t,\lambda_t)p(x_t,t;\pspeed)\}'/p(x_{t'},t')\right]}
  \end{align}
  with
  \begin{equation}
    \label{eq:derivation_relation_x_dot_J_t}
    J(x_t,t) \equiv d(x_t,\lambda_t) j(x_t,t)/p(x_t,t) + D \{d(x_t,\lambda_t)p(x_t,t)\}'/p(x_t,t),
  \end{equation}
  where
  \begin{equation}
    \label{eq:derivation_relation_x_dot_derivative}
    \{d(x_t,\lambda_t)p(x_t,t)\}' \equiv \partial_x
    d(x,\lambda_t)p(x,t)\vert_{x = x_t}.
  \end{equation}
  We introduce the shorthand notation $C_t\equiv
  C(x_t,\lambda_t)$ for an arbitrary state function $C(x_t,\lambda_t)$. First, we use the Langevin
  eq.~\eqref{eq:setup_langevin_eq} to rewrite the expression
  \begin{equation}
    \label{eq:appendix_correlation_functions_noise_terms_1}
    \expval{d_t \circ\dot{x}_t
      d_{t'}\circ\dot{x}_{t'}} = \expval{d_t\circ\zeta_t
      d_{t'}\circ\zeta_{t'}} - \expval{d_t \mu \Ftot_t d_{t'}\mu \Ftot_{t'}} +
    \expval{d_t\mu \Ftot_t d_{t'}\circ \dot{x}_{t'}} + \expval{d_{t'}\mu \Ftot_{t'} d_{t}\circ \dot{x}_{t}}
  \end{equation}
  in terms of the noise. Then, we use Itô's lemma~\cite{gardiner} to write the
  first term in eq.~\eqref{eq:appendix_correlation_functions_noise_terms_1} in
  terms of non-anticipating functions and the noise, i.e., 
  \begin{equation}
    \label{eq:appendix_correlation_functions_noise_terms_ito_lemma_first_term}
    \expval{d_t\circ\zeta_t d_{t'}\circ\zeta_{t'}} = \expval{d_{t}\cdot\zeta_t
    d_{t'}\cdot\zeta_{t'}} - D^2\expval{d'_{t}d'_{t'}} +
  \expval{Dd'_{t}d_{t'}\circ\left[\dot{x}_{t'}-\mu\Ftot_{t'}\right]} + \expval{Dd'_{t'}d_{t}\circ\left[\dot{x}_{t}-\mu\Ftot_{t}\right]},
  \end{equation}
  where $\cdot$ denotes the Itô product and 
  \begin{equation}
    \label{eq:appendix_correlation_functions_derivative_of_x}
    d'_t \equiv \partial_x d(x,\lambda_t)\vert_{x=x_t}.
  \end{equation}
  Next, we use Itô's Isometry~\cite{gardiner} to evaluate the first term in
  eq.~\eqref{eq:appendix_correlation_functions_noise_terms_ito_lemma_first_term},
  which reads
  \begin{equation}
    \label{eq:appendix_correlation_functions_first_term_ito_isometry}
    \expval{d_{t}\cdot\zeta_t d_{t'}\cdot\zeta_{t'}} = 2D \delta(t-t')\expval{d^2_t}.
  \end{equation}
  Using
  eqs.~\eqref{eq:appendix_correlation_functions_noise_terms_ito_lemma_first_term}
  and~\eqref{eq:appendix_correlation_functions_first_term_ito_isometry} we can
  rewrite eq.~\eqref{eq:appendix_correlation_functions_noise_terms_1} as
  \begin{align}
    \label{eq:appendix_correlation_functions_term_evaluated_with_delta}
    \expval{d_t \circ\dot{x}_t d_{t'}\circ\dot{x}_{t'}} =&
    2D\delta(t-t')\expval{d^2_t} - \expval{\left[d_{t}\mu\Ftot_{t} + D
d'_{t}\right]\left[d_{t'}\mu\Ftot_{t'} + D d'_{t'}\right]} \\
    &+\expval{\left[d_{t'}\mu\Ftot_{t'} + D d'_{t'}\right] d_{t}\circ\dot{x}_{t}}
      +\expval{\left[d_{t}\mu\Ftot_{t} + D d'_{t}\right] d_{t'}\circ\dot{x}_{t'}}.\nonumber
  \end{align}

  Now, we rewrite the last two terms of
  eq.~\eqref{eq:appendix_correlation_functions_term_evaluated_with_delta} in
  terms of correlation functions between state functions. To do so, we define
  a variable $B(x,\lambda_t)$ such that $\partial_x B(x,\lambda_t) =
  d(x,\lambda_t)$ and hence, $d(x_t,\lambda_t)\circ \dot{x}_t =
  \dot{B}(x_t,\lambda_t) - \dot{\lambda}\partial_\lambda
  B(x,\lambda)\vert_{\lambda=\lambda_t}$, where $\dot{B}(x_t,\lambda_t)\equiv (\mathrm{d}/\mathrm{d}t)B(x_t,\lambda_t)$ is the total
  time-derivative of the state function $B(x,\lambda_t)$. The last term
  in eq.~\eqref{eq:appendix_correlation_functions_term_evaluated_with_delta}
  can be written as
  \begin{equation}
    \label{eq:appendix_correlation_functions_change_notation_derivative_correlation}
    \expval{\left[d_{t}\mu\Ftot_{t} + D d'_{t}\right] d_{t'}\circ\dot{x}_{t'}}
    = \expval{A_t \circ\left[\dot{B}_{t'} - \dot{\lambda}\partial_\lambda B(x_{t'},\lambda)\vert_{\lambda=\lambda_{t'}}\right]},
  \end{equation}
  with $A_t \equiv d_{t}\mu\Ftot_{t} + D d'_{t}$. Next, we can rewrite the
  first term on the r.h.s of
  eq.~\eqref{eq:appendix_correlation_functions_change_notation_derivative_correlation}
  by applying the derivative after averaging because average values and time
  derivatives commute in the Stratonovich convention~\cite{zinn},
  i.e.,
  \begin{equation}
    \label{eq:appendix_correlation_functions_time_derivative_commute_t_t'}
    \expval{A_t \dot{B}_{t'}} = (\mathrm{d}/\mathrm{d}t') \expval{A_t B_{t'}} =
    (\mathrm{d}/\mathrm{d}t')\int\dd{x} A(x,\lambda_t) \TEOunscaled{x}{t}{t'} B(x,\lambda_{t'})
  \end{equation}
  for $t>t'$ and 
  \begin{equation}
    \label{eq:appendix_correlation_functions_time_derivative_commute_t'_t}
    \expval{A_t \dot{B}_{t'}} =
    (\mathrm{d}/\mathrm{d}t')\int\dd{x} B(x,\lambda_{t'}) \TEOunscaled{x}{t'}{t} A(x,\lambda_{t})
  \end{equation}
  for $t'>t$ with time evolution operator
  \begin{equation}
    \label{eq:appendix_correlation_functions_time_evolution_operator}
    \TEOunscaled{x}{t'}{t}\equiv \TOexp{\int_t^{t'}\dd{t''}\FPO[x,\lambda_{t''}]}
  \end{equation}
  and Fokker-Planck operator
  \begin{equation}
    \label{eq:appendix_correlation_functions_fokker_planck_operator}
    \FPO \equiv -\partial_x\left(\mobility\Ftot - \D\partial_x\right).
  \end{equation}
 Using
  $(\mathrm{d}/\mathrm{d}t')\TEOunscaled{x}{t'}{t}=\FPO[x,t']\TEOunscaled{x}{t'}{t}$ and
  $(\mathrm{d}/\mathrm{d}t')\TEOunscaled{x}{t}{t'}=\TEOunscaled{x}{t}{t'}\FPO[x,t']$ for eqs.~\eqref{eq:appendix_correlation_functions_time_derivative_commute_t_t'}
and~\eqref{eq:appendix_correlation_functions_time_derivative_commute_t'_t} yields the following expression 
\begin{align}
  \label{eq:appendix_correlation_functions_correlation_between_state_function}
    \expval{A_t \dot{B}_{t'}} &= \theta(t-t') \expval{A_t\left[2\nu_{t'} d_{t'} -
      \mu\Ftot_{t'} - D d'_{t'} + \dot{\lambda}\partial_\lambda
                                  B(x_{t'},\lambda)\vert_{\lambda=\lambda_{t'}}\right]}\\
                                &+\theta(t'-t)\expval{A_t\left[\mu\Ftot_{t'}d_{t'}
                                  + D d'_{t'} + \dot{\lambda}\partial_\lambda
                                  B(x_{t'},\lambda)\vert_{\lambda=\lambda_{t'}}\right]},\nonumber
\end{align}
where $\theta(\cdot)$ is the Heaviside function and $\nu_t \equiv
j(x_t,\lambda_t)/p(x_t,\lambda_t)$ is the mean local velocity. Finally,
inserting
eq.~\eqref{eq:appendix_correlation_functions_correlation_between_state_function}
into eq.~\eqref{eq:appendix_correlation_functions_term_evaluated_with_delta}
leads to
\begin{align}
  \label{eq:appendix_correlation_functions_final_term}
  \expval{d_t \circ\dot{x}_t d_{t'}\circ\dot{x}_{t'}} &=
  2D\delta(t-t')\expval{d^2_t} \\
  &+ \theta(t-t') \expval{\left[\mu\Ftot_{t} d_{t}
      + Dd'_{t}\right]\left[2\nu_{t'}d_{t'}-\mu\Ftot_{t'}d_{t'}-Dd'_{t'}\right]}\nonumber\\
  &+ \theta(t'-t) \expval{\left[\mu\Ftot_{t'} d_{t'} + Dd'_{t'}\right]\left[2\nu_{t}d_{t}-\mu\Ftot_{t}d_{t}-Dd'_{t}\right]},\nonumber
\end{align}
which is identical to eq.~\eqref{eq:derivation_relation_x_dot} when
identifying the terms $J(x_t,\lambda_t)=\mu\Ftot_{t'} d_{t'} + Dd'_{t'}$ and
$J(x_t,\lambda_t) - 2D{d(x_t,\lambda_t)p(x_t,\lambda_t)}'/p(x_t,t) =2\nu_{t}d_{t}-\mu\Ftot_{t}d_{t}-Dd'_{t}$.

  \section{Limit of Fast Driving}
  \label{sec:appendix_limit_of_fast_driving}
  In this section, we derive the scaling properties of the quality factors in
  the limit of fast driving shown in table~\ref{tab:ATDD_Fast_Driving}.

  First, we determine the leading order of the total entropy production rate by
  inserting eqs.~\eqref{eq:derivation_fast_driving_density} and~\eqref{eq:derivation_fast_driving_probabilty_current} into
  eq.~\eqref{eq:derivation_scaled_total_entropy_production_rate}, which yields
  \begin{equation}
    \label{eq:derivation_fast_driving_total_entropy_production_rate}
    \sigma (\T,\pspeed) = \frac{1}{\tsys}\sigma^{(0)}(\T,\pspeed)+ \order{\epsf/\tsys} \equiv\frac{1}{\tsys\tf}\int_0^{\tf}\dd{\tau} \int \dd{x}
    \frac{\jorder{0}(x,\tau)^2}{\sD p(x,0)} + \order{\epsf/\tsys}.
  \end{equation}
  Obviously, the entropy production rate is of $\order{1}$ in
  the limit of fast driving. 

  Next, we determine the leading orders of the mean values and their response terms.
  Inserting eqs.~\eqref{eq:derivation_fast_driving_density},~\eqref{eq:derivation_fast_driving_density_zero_order},~\eqref{eq:derivation_fast_driving_density_first_order}
  into~\eqref{eq:derivation_mean_value_instant_state_var} leads to an expression
  for the instantaneous state variable
  \begin{equation}
    \label{eq:derivation_fast_driving_mean_state_var}
    \Aim = \Aimorder{0} + \epsf \Aimorder{1} + \order{\epsf^2}
  \end{equation}
  with
  \begin{equation}
    \label{eq:derivation_fast_driving_mean_state_var_zero_order}
    \Aimorder{0} \equiv \int\dd{x} a(x,\lambda_{\tf})p(x,0)
  \end{equation}
  and
  \begin{equation}
    \label{eq:derivation_fast_driving_mean_state_var_first_order}
    \Aimorder{1} \equiv \int\dd{x} a(x,\lambda_{\tf}) \FPOeff[x,\tf,0] p(x,0).
  \end{equation}
  The response term of $\Aim$ is consequently given by
  \begin{equation}
    \label{eq:derivation_fast_driving_response_mean_state_var}
    \response{a} \equiv \left[\DDelta \Aim\right]^2 = \left[\Aimorder{1}\right]^2\epsf^2 + \order{\epsf^3}
  \end{equation}
  due to the fact that $\DDelta f(\tf=\pspeed\T) = 0$ for an arbitrary
  function depending only on $\tf$ and not depending separately on $\pspeed$ and
  $\T$. For time-averaged state variables~\eqref{eq:derivation_mean_value_time_averaged_state_var} one gets a
  similar behavior by following the analogous steps above, which leads to the
  response term
  \begin{equation}
    \label{eq:derivation_fast_driving_response_mean_time_averaged_state_var}
    \response{A} \equiv \left[\DDelta \Acm\right]^2 = \left[\Acmorder{1}\right]^2\epsf^2 + \order{\epsf^3}
  \end{equation}
  with
  \begin{equation}
    \label{eq:derivation_fast_driving_mean_time_averaged_state_var_first_order}
    \Acmorder{1} \equiv \frac{1}{\tf}\int_0^{\tf}\int\dd{x} a(x,\lambda_{\tau}) \FPOeff[x,\tau,0] p(x,0).
  \end{equation}
  Furthermore, for the current defined
  in eq.~\eqref{eq:derivation_mean_value_current_I} one finds the following
  expression for the current
  \begin{equation}
    \label{eq:derivation_fast_driving_mean_current_I}
    \Jtm{b} = \frac{1}{\epsf\tsys}\left(\Jtmorder{0}{b} + \epsf \Jtmorder{1}{b} + \order{\epsf^2}\right)  
  \end{equation}
  with zeroth order
  \begin{equation}
    \label{eq:derivation_fast_driving_current_I_zero_order}
    \Jtmorder{0}{b} \equiv \frac{1}{\tf}\int_0^{\tf}\int\dd{x} \dot{b}(x,\lambda_{\tau}) p(x,0)
  \end{equation}
  and first order
  \begin{equation}
    \label{eq:derivation_fast_driving_current_II_first_order}
    \Jtmorder{1}{b} \equiv \frac{1}{\tf}\int_0^{\tf}\int\dd{x}
    \dot{b}(x,\lambda_{\tau}) \FPOeff[x,\tau,0]p(x,0).
  \end{equation}
  Using these expressions leads to the response term
  \begin{equation}
    \label{eq:derivation_fast_driving_response_mean_current_I}
    \response{J_b} \equiv \left[\Jtm{b} + \DDelta \Jtm{b}\right]^2 =
    \left[\Jtmorder{1}{b}\right]^2 (1/\tsys^2) + \order{\epsf}.
  \end{equation}
  Moreover, for the current depending on the velocity we insert
  eqs.~\eqref{eq:derivation_fast_driving_probabilty_current},~\eqref{eq:derivation_fast_driving_current_zero_order}
  into eq.~\eqref{eq:derivation_probability_current}, which leads to
  \begin{equation}
    \label{eq:derivation_fast_driving_mean_current_II}
    \Jtm{d} = \frac{1}{\tsys} \Jtmorder{0}{d} + \order{\epsf/\tsys}
  \end{equation}
  with zeroth leading order
  \begin{equation}
    \label{eq:derivation_fast_driving_mean_current_II_zero_order}
    \Jtmorder{0}{d} \equiv \frac{1}{\tf}\int_0^{\tf}\dd{\tau}\int\dd{x}
    d(x,\lambda_\tau) \jorder{0}(x,\tau).
  \end{equation}
  The response term consequently reads
  \begin{equation}
    \label{eq:derivation_fast_driving_response_mean_current_II}
    \response{J_d} \equiv \left[\Jtm{d} + \DDelta \Jtm{d}\right]^2 =
    \left[\Jtmorder{0}{d}\right]^2 \frac{1}{\tsys^2} + \order{\epsf/\tsys^2}.
  \end{equation}

  Now we derive the leading order of the diffusion coefficients.
  First, using the leading order of the density
  in~\eqref{eq:derivation_fast_driving_density_zero_order} and plugging it into
  eq.~\eqref{eq:derivation_diff_coeff_value_instant_state_var} yields
  \begin{equation}
    \label{eq:derivation_fast_driving_diff_coeff_state_var}
    D_a(\T,\pspeed) = \tsys\tf\epsf D^{(1)}_a(\T,\pspeed) + \order{\epsf^2}
  \end{equation}
  with
  \begin{equation}
    \label{eq:derivation_fast_driving_diff_coeff_state_var_zero_order}
    D^{(1)}_a(\T,\pspeed) \equiv \frac{1}{2}\left(\int\dd{x} a(x,\lambda_{\tf})^2p(x,0) - \left[\int\dd{x} a(x,\lambda_{\tf})p(x,0)\right]^2\right)
  \end{equation}
  for the instantaneous state variable.
  For all other diffusion coefficients it is sufficient to use the zeroth order
  of the propagator defined in
  eq.~\eqref{eq:derivation_fast_driving_propagator_zero_order}. Plugging this
  leading order into the diffusion
  coefficient~\eqref{eq:derivation_diff_coeff_value_time_averaged_state_var} for
  time-averaged state variable leads to
  \begin{equation}
    \label{eq:derivation_fast_driving_diff_coeff_time_averaged_state_var}
    D_A(\T,\pspeed) = \tsys\tf\epsf D^{(1)}_A(\T,\pspeed) + \order{\epsf^2}
  \end{equation}
  with
  \begin{equation}
    \label{eq:derivation_fast_driving_diff_coeff_time_averaged_state_var_zero_order}
    D^{(1)}_A(\T,\pspeed) \equiv
    \frac{1}{\tf^2}\int_0^{\tf}\dd{\tau}\int_0^\tau\dd{\tau'}\int\dd{x}
    a(x,\lambda_{\tau})a(x,\lambda_{\tau'})p(x,0) - \frac{1}{2}\Acmorder{0}^2,
  \end{equation}
  where
  \begin{equation}
    \label{eq:derivation_fast_driving_diff_coeff_time_averaged_state_var_mean_zero}
    \Acmorder{0} \equiv \frac{1}{\tf}\int_0^{\tf}\dd{\tau}\int\dd{x}a(x,\lambda_\tau)p(x,0)
  \end{equation}
  is the leading zeroth order of the time-averaged state variable.
  Furthermore, plugging
  eq.~\eqref{eq:derivation_fast_driving_propagator_zero_order}
  into~\eqref{eq:derivation_mean_value_current_I} yields the diffusion
  coefficient
  \begin{equation}
    \label{eq:derivation_fast_driving_diff_coeff_current_I}
    D_{J_b} = \frac{\tf}{\tsys\epsf}\left(D^{(0)}_{J_b}(\T, \pspeed) + \order{\epsf}\right)
  \end{equation}
  of the current $\Jtm{b}$ with
  \begin{equation}
    \label{eq:derivation_fast_driving_diff_coeff_current_I_zero_order}
    D^{(0)}_{J_b}(\T, \pspeed) \equiv
    \frac{1}{\tf^2}\int_0^{\tf}\dd{\tau}\int_0^\tau\dd{\tau'}\int\dd{x}
    \dot{b}(x,\lambda_{\tau})\dot{b}(x,\lambda_{\tau'})p(x,0) - \frac{1}{2}\Jtmorder{0}{b}^2,
  \end{equation}
  where $\Jtmorder{0}{b}$ is defined in
  eq.~\eqref{eq:derivation_fast_driving_current_I_zero_order}.
  For diffusion coefficient of the current depending on the
  velocity, we insert eq.~\eqref{eq:derivation_fast_driving_propagator_zero_order}
  into~\eqref{eq:derivation_diffusion_coefficient_current_II}, which leads to
  \begin{equation}
    \label{eq:derivation_fast_driving_diff_coeff_current_II}
    D_{J_d}(\T,\pspeed) = \frac{1}{\tsys}D^{(0)}_{J_d}(\T,\pspeed) + \order{\epsf/\tsys}
  \end{equation}
  with leading order
  \begin{equation}
    \label{eq:derivation_fast_driving_diff_coeff_current_II_zero_order}
    D^{(0)}_{J_d}(\T,\pspeed) \equiv \frac{1}{\tf}\int_0^{\tf}\dd{\tau}\int\dd{x}
    d^2(x,\lambda_\tau) p(x,0).
  \end{equation}

  Finally, we use the above derived results to determine the leading orders of
  all quality factors.
  First, by using eqs.~\eqref{eq:derivation_fast_driving_total_entropy_production_rate},
  ~\eqref{eq:derivation_fast_driving_response_mean_state_var}
  and~\eqref{eq:derivation_fast_driving_diff_coeff_state_var} we can determine
  the quality factor
  \begin{equation}
    \label{eq:derivation_fast_driving_quality_factor_instant_state_var}
    Q_a \equiv \frac{\response{a}}{\sigma(\T,\pspeed)D_a(\T,\pspeed)} \approx
    \frac{\left[\Aimorder{1}\right]^2}{\sigma^{(0)}(\T,\pspeed)\tf D^{(1)}_a(\T,\pspeed) }\epsf
  \end{equation}
  for the instantaneous state variable $\Aim$. Furthermore, via
  eqs.~\eqref{eq:derivation_fast_driving_total_entropy_production_rate},
  \eqref{eq:derivation_fast_driving_response_mean_time_averaged_state_var}
  and~\eqref{eq:derivation_fast_driving_diff_coeff_time_averaged_state_var} we
  get the asymptotic behavior of the quality factor
  \begin{equation}
    \label{eq:derivation_fast_driving_quality_factor_time_averaged_state_var}
    Q_A \equiv \frac{\response{A}}{\sigma(\T,\pspeed)D_A(\T,\pspeed)} \approx
    \frac{\left[\Acmorder{1}\right]^2}{\sigma^{(0)}(\T,\pspeed)\tf D^{(1)}_A(\T,\pspeed)}\epsf
  \end{equation}
  for the time-averaged observable. Moreover using
  eqs.~\eqref{eq:derivation_fast_driving_total_entropy_production_rate},
  ~\eqref{eq:derivation_fast_driving_response_mean_current_I}
  and~\eqref{eq:derivation_fast_driving_diff_coeff_current_I} yields the quality
  factor
  \begin{equation}
    \label{eq:derivation_fast_driving_quality_factor_current_I}
    Q_{J_b} \equiv \frac{\response{J_b}}{\sigma(\T,\pspeed)D_{J_b}(\T,\pspeed)}
    \approx \frac{\left[\Jtmorder{1}{b}\right]^2}{\sigma^{(0)}(\T,\pspeed)\tf
      D^{(0)}_{J_b}(\T,\pspeed)} \epsf
  \end{equation}
  for the current depending on the time spent in a certain state.
  Last but not least, using
  eqs.~\eqref{eq:derivation_fast_driving_total_entropy_production_rate},
  ~\eqref{eq:derivation_fast_driving_response_mean_current_II}
  and~\eqref{eq:derivation_fast_driving_diff_coeff_current_II} leads to the
  expression for the quality factor
  \begin{equation}
    \label{eq:derivation_fast_driving_quality_factor_current_II}
    Q_{J_d} \equiv \frac{\response{J_d}}{\sigma(\T,\pspeed) D_{J_d}(\T,\pspeed)}
    \approx \frac{\left[\Jtmorder{0}{d}\right]^2}{\sigma^{(0)}(\T,\pspeed) D^{(0)}(\T,\pspeed)}
  \end{equation}
  for the current depending on the velocity. We remark, that the explicit expression for the quality
  factor~\eqref{eq:derivation_fast_driving_quality_factor_current_II} is given
  by eq.~\eqref{eq:derivation_fast_driving_quality_factor_II_explicit} in the main
  text, which can be verified by using
  eqs.~\eqref{eq:derivation_fast_driving_total_entropy_production_rate},~\eqref{eq:derivation_fast_driving_mean_current_II}
  and~\eqref{eq:derivation_fast_driving_diff_coeff_current_II}.

  \section{Limit of Slow Driving}
  \label{sec:appendix_limit_of_slow_driving}
  In this section, we derive the generic scaling properties of the quality
  factors in the limit of slow driving shown in
  table~\eqref{tab:ATDD_Slow_Driving}. A system prepared in an arbitrary
  initial condition relaxes into the stationary state at $\lambda_0$ on a
  time scale that is much shorter than the time scale of the external
  driving on which the protocol changes.
  We are interested in the dynamics on a time scale that is comparable
  with the time scale of the driving and hence, we assume that the
  system has already reached the stationary state at $\lambda_0$.
  
  We first derive an expression for the total entropy production rate. For
  this, we insert
  eqs.~\eqref{eq:derivation_slow_driving_density}
  and~\eqref{eq:derivation_slow_driving_probability_current} into
  eq.~\eqref{eq:derivation_scaled_total_entropy_production_rate}, which leads to
  \begin{equation}
    \label{eq:derivation_slow_driving_total_entropy_production_rate}
    \sigma(\T,\pspeed) =
    \frac{1}{\tsys\tf}\int_0^{\tf}\dd{\tau}\int\dd{x}\frac{\left(\jorder{0}(x,\tau)
        + \epss \jorder{1}(x,\tau)\right)^2}{\sD\porder{0}(x,\tau)}.
  \end{equation}
  If the system is driven around an equilibrium state, the entropy production
  rate vanishes, i.e., $\sigma(\T,\pspeed)=\order{\epss}$. If the system is
  driven around a NESS, the entropy production rate is finite and hence,
  $\sigma(\T,\pspeed)=\order{1}$.

  Next, we derive the leading orders of the mean values and their response terms.
  Due to the fact, that the
  density~\eqref{eq:derivation_slow_driving_density_zero_order} is a function of
  the protocol in the leading order, the response term of the zeroth order of
  the instantaneous state variable vanishes. As a consequence, the response term
  of this quantity is given by
  \begin{equation}
    \label{eq:derivation_slow_driving_response_instantaneous_state_variable}
    \response{a} = \left[\Aimorder{1}\right]^2\epss^2 + \order{\epss^3},
  \end{equation}
  where we used that $\DDelta \epss = -\epss$.
  This implies that the response terms for the time-averaged state variable
  \begin{equation}
    \label{eq:derivation_slow_driving_response_time_avergaed_state_variable}
    \response{A} = \left[\Acmorder{1}\right]^2\epss^2 + \order{\epss^3}
  \end{equation}
  as well as for the current depending on the residence time
  \begin{equation}
    \label{eq:derivation_slow_driving_response_current_residence_time}
    \response{J_b} = \left[\Jtmorder{2}{b}\right]^2\epss^4/\tsys^2 + \order{\epss^5}
  \end{equation}
  vanish asymptotically.
  The response term for the current depending on the velocity is given by
  \begin{equation}
    \label{eq:derivation_slow_driving_response_current_velocity}
    \response{J_d} = \left[\Jtmorder{0}{d} - \epss^2 \Jtmorder{2}{d} + \order{\epss^3}\right]^2
    (1/\tsys)^2,
  \end{equation}
  where the linear term in $\epss$ of the current vanishes due to $\DDelta
  \epss= -\epss$. Depending on whether a non-conservative force is applied or
  not, the response
  term~\eqref{eq:derivation_slow_driving_response_current_velocity} is either
  of $\order{1}$ or $\order{\epss^4}$, respectively.

  Now, we derive the leading orders of the diffusion coefficients. First, for
  the instantaneous state variable we plug
  eq.~\eqref{eq:derivation_slow_driving_density}
  into~\eqref{eq:derivation_diff_coeff_value_instant_state_var} and obtain
  \begin{equation}
    \label{eq:appendix_slow_driving_diffusion_coefficient_instantaneous_state_variable}
    D_a(\T,\pspeed) = \frac{\tau_f\tsys}{\epss}\left[D^{(0)}_a(\T,\pspeed) + \order{\epss}\right]
  \end{equation}
  with
  \begin{equation}
    \label{eq:appendix_slow_driving_diffusion_coefficient_instantaneous_state_variable_zeroth_order}
    D^{(0)}_a(\T,\pspeed)\equiv \frac{1}{2}\int\dd{x}a^2(x,\lambda_{\tf})\porder{0}(x,\tf)-\left[\int\dd{x}a(x,\tf)\porder{0}(x,\tf)\right]^2.
  \end{equation}
  For the time-averaged observables, we
  insert the propagator~\eqref{eq:derivation_slow_driving_propagator} in the
  limit of slow driving into the diffusion
  coefficients~\eqref{eq:derivation_diff_coeff_value_time_averaged_state_var},~\eqref{eq:derivation_diffusion_coefficient_current_I}
  and~\eqref{eq:derivation_diffusion_coefficient_current_II}. The leading
  order of the variance of the time-averaged state variable vanishes such that
  its diffusion coefficient is given by
  \begin{equation}
    \label{eq:appendix_slow_driving_diffusion_coefficient_time_averaged_state_variable}
    D_A(\T,\pspeed) = \tf\tsys D_A^{(0)}(\T,\pspeed) + \order{\epss}
  \end{equation}
  with leading order $D_A^{(0)}(\T,\pspeed)$. Analogously, for the
  current $J_b(\T,\pspeed)$ we find
  \begin{equation}
    \label{eq:appendix_slow_driving_diffusion_coefficient_current_I}
    D_{J_b}(\T,\pspeed) = \frac{\epss^2}{\tf\tsys} D^{(2)}_{J_b}(\T,\pspeed) + \order{\epss^3}
  \end{equation}
  with leading order $D^{(2)}_{J_b}(\T,\pspeed)$. Furthermore, the diffusion
  coefficient for the current $J_d(\T,\pspeed)$ converges to
  \begin{equation}
    \label{eq:appendix_slow_driving_diffusion_coefficient_current_II}
    D_{J_d}(\T,\pspeed) = \frac{1}{\tsys} D^{(0)}_{J_d}(\T,\pspeed) + \order{\epss}
  \end{equation}
  with leading order $D^{(0)}_{J_d}(\T,\pspeed)$ as the last two terms in
  eq.~\eqref{eq:derivation_diffusion_coefficient_current_II} compensate each other, when
  using~\eqref{eq:derivation_slow_driving_propagator}.

  Lastly, we determine the leading orders of all quality factors by using the
  above derived results. Using
  eqs.~\eqref{eq:derivation_slow_driving_total_entropy_production_rate},~\eqref{eq:derivation_slow_driving_response_instantaneous_state_variable},
  and~\eqref{eq:appendix_slow_driving_diffusion_coefficient_instantaneous_state_variable}
  leads to the quality factor
  \begin{equation}
    \label{eq:appendix_slow_driving_quality_factor_instantaneous_state_variable}
    \mathcal{Q}_a \equiv
    \frac{\response{a}}{\sigma(\T,\pspeed)D_a(\T,\pspeed)}\approx
    \frac{\left[\Aimorder{1}\right]^2\epss^3}{D_a^{(0)}(\T,\pspeed)\int_0^{\tf}\dd{\tau}\int\dd{x}\frac{\left(\jorder{0}(x,\tau)
        + \epss \jorder{1}(x,\tau)\right)^2}{\sD\porder{0}(x,\tau)}}
  \end{equation}
  of the instantaneous state variable. Depending on whether a non-conservative
  force is applied or not the quality
  factor~\eqref{eq:appendix_slow_driving_quality_factor_instantaneous_state_variable}
  vanishes like $\epss^3$ or $\epss$, respectively. The quality factor of the
  time-averaged state variable can be calculated by using
  eqs.~\eqref{eq:derivation_slow_driving_total_entropy_production_rate},~\eqref{eq:derivation_slow_driving_response_time_avergaed_state_variable}
  and~\eqref{eq:appendix_slow_driving_diffusion_coefficient_time_averaged_state_variable}
  and reads
  \begin{equation}
    \label{eq:appendix_slow_driving_quality_factor_time_averaged_state_variable}
    \mathcal{Q}_A \equiv\frac{\response{A}}{\sigma(\T,\pspeed)D_A(\T,\pspeed)}\approx\frac{\left[\Acmorder{1}\right]^2\epss^2}{D^{(0)}_A(\T,\pspeed)\int_0^{\tf}\dd{\tau}\int\dd{x}\frac{\left(\jorder{0}(x,\tau)
          + \epss \jorder{1}(x,\tau)\right)^2}{\sD\porder{0}(x,\tau)}}.
  \end{equation}
  This quality factor is either of $\order{\epss^2}$ or $\order{1}$ depending on
  whether the system is in a NESS or in an equilibrium state at fixed
  $\lambda_\tau$. Furthermore, we use
  eqs.~\eqref{eq:derivation_slow_driving_total_entropy_production_rate},~\eqref{eq:derivation_slow_driving_response_current_residence_time}
  and~\eqref{eq:appendix_slow_driving_diffusion_coefficient_current_I} to
  determine the leading order of the quality factor for current
  $J_b(\T,\pspeed)$
  \begin{equation}
    \label{eq:appendix_slow_driving_quality_factor_current_I}
    \mathcal{Q}_{J_b} \equiv \frac{\response{J_b}}{\sigma(\T,\pspeed)D_{J_b}(\T,\pspeed)}\approx\frac{\left[\Jtmorder{2}{b}\right]^2\epss^2\tf^2}{D_{J_b}(\T,\pspeed)\int_0^{\tf}\dd{\tau}\int\dd{x}\frac{\left(\jorder{0}(x,\tau)
          + \epss \jorder{1}(x,\tau)\right)^2}{\sD\porder{0}(x,\tau)}},
  \end{equation}
  which vanishes like $\epss^2$, if a non-conservative force is applied and is
  of $\order{1}$, when only conservative forces are applied. Using
  eqs.~\eqref{eq:derivation_slow_driving_total_entropy_production_rate},~\eqref{eq:derivation_slow_driving_response_current_velocity}
  and~\eqref{eq:appendix_slow_driving_diffusion_coefficient_current_II} yields
  the quality factor for the current $\Jtm{d}$
  \begin{equation}
    \label{eq:appendix_slow_driving_quality_factor_current_II}
    \mathcal{Q}_{J_d}\equiv\frac{\response{J_d}}{\sigma(\T,\pspeed)D_{J_d}(\T,\pspeed)}\approx\frac{\left[\Jtmorder{0}{d}
      -\epss^2\Jtmorder{2}{d}\right]^2\tf}{D^{(0)}_{J_d}(\T,\pspeed)\int_0^{\tf}\dd{\tau}\int\dd{x}\frac{\left(\jorder{0}(x,\tau)
          + \epss \jorder{1}(x,\tau)\right)^2}{\sD\porder{0}(x,\tau)}}.
  \end{equation}
  The quality
  factor~\eqref{eq:appendix_slow_driving_quality_factor_current_II} is of
  $\order{1}$, if a non-conservative force is applied and vanishes like
  $\epss^2$, if only conservative forces are present.

\end{appendix}

\section*{References}
\bibliographystyle{iopart-num}
\bibliography{/Users/timikoyuk/work/GitRepositories/Bibliography/refs.bib}

\end{document}